\documentclass[12pt]{article}
\usepackage[utf8]{inputenc}
\usepackage{authblk}

\title{\vspace{-2.0cm}\large An assessment of racial disparities in pretrial decision-making using misclassification models}

\author[1]{Kimberly A. Hochstedler Webb}
\author[2]{Sarah A. Riley}
\author[1]{Martin T. Wells}
\affil[1]{Department of Statistics and Data Science, Cornell University}
\affil[2]{Department of Communication, Stanford University}

\date{}

\usepackage{natbib}
\usepackage{graphicx}
\usepackage[paperwidth=8.5in,paperheight=11.0in,top=1in, bottom=1in, left=1in, right=1in,lines=25]{geometry}
\usepackage[title,toc,titletoc,page]{appendix}

\usepackage{siunitx}
\usepackage{appendix}
\usepackage{mathtools}
\usepackage{threeparttable}
\usepackage{bbm}
\usepackage{caption}
\usepackage{algpseudocode}
\usepackage{algorithm}
\usepackage{url}

\usepackage{ccaption}
\usepackage{booktabs}

\usepackage{setspace}
\usepackage{stix}

\begin{document}

\maketitle

\begin{abstract}
Pretrial risk assessment tools are used in jurisdictions across the country to assess the likelihood of ``pretrial failure,'' the event where defendants either fail to appear for court or reoffend. Judicial officers, in turn, use these assessments to determine whether to release or detain defendants during trial. While algorithmic risk assessment tools were designed to predict pretrial failure with greater accuracy relative to judges, there is still concern that both risk assessment recommendations and pretrial decisions are biased against minority groups. In this paper, we develop methods to investigate the association between risk factors and pretrial failure, while simultaneously estimating misclassification rates of pretrial risk assessments and of judicial decisions as a function of defendant race. This approach adds to a growing literature that makes use of outcome misclassification methods to answer questions about fairness in pretrial decision-making. We give a detailed simulation study for our proposed methodology and apply these methods to data from the Virginia Department of Criminal Justice Services. We estimate that the VPRAI algorithm has near-perfect specificity, but its sensitivity differs by defendant race. Judicial decisions also display evidence of bias; we estimate wrongful detention rates of $39.7\%$ and $51.4\%$ among white and Black defendants, respectively. 

\end{abstract}

\newpage

\section{Introduction}
Algorithmic pretrial risk assessments provide decision-makers with an assessment of the likelihood of ``pretrial failure,'' the event where defendants either fail to appear for court (FTA) or reoffend. Judges use them at arraignment to determine whether to release or detain defendants pending trial. In theory, by predicting pretrial failure with greater accuracy and objectivity relative to judges, risk assessments simultaneously promote decarceration, public safety, and racial equity \citep{milgram2015pretrial, viljoen2019impact, marlowe2020employing}. In doing so, they solve multiple policy problems at once: reducing costs and alleviating overburdened carceral infrastructure, keeping the public safe, and mitigating long standing inequality in the criminal legal system. 

Jurisdictions across the country have taken advantage of this low-cost policy solution, and algorithmic pretrial risk assessments are rapidly gaining popularity. Between 2014 and 2017, the number of US jurisdictions using pretrial risk assessments increased twofold \citep{TheStateofPretrialJusticeinAmerica_2017}, so that today roughly half report using them \citep{ScanofPretrialPractices_2019}, and two thirds of the population is subject to their use \citep{lattimore2020prevalence}. The Virginia Pretrial Risk Assessment Instrument (VPRAI) alone is used across the entire state of Virginia and 59 counties outside the state, so that an estimated 34.2 million people live in a county that administers that particular risk tool \citep{HowManyJurisdictionsUseEachTool?_2023}.

The evidence as to whether or not risk assessments achieve the desired policy goals is mixed, in part because the body of empirical research documenting their impacts is still relatively small \citep{copp2022pretrial}. Theoretical work comparing pretrial risk assessment predictions to judicial decisions suggests that pretrial risk assessments may lead to reductions in crime, decreased jail populations, and greater racial equity \citep{kleinberg2018human}. Even among subgroups of defendants, pretrial risk assessment tools have been shown to effectively predict subsequent recidivism \citep{cohen2018predicting}. However, as many have pointed out \citep{cadigan2011implementing, stevenson2018assessing, stevenson2022algorithmic, copp2022pretrial}, judges—not risk instruments—make pretrial decisions, and understanding the effects of pretrial risk assessments in terms of jail populations, racial disparities, and criminal activity requires empirical study. The results of multiple quasi-experimental studies suggest that pretrial risk assessments do not fulfill their potential. In some cases, the initial reductions in pretrial detention diminish after only a few months \citep{sloan2023effect}, and in others they do not appear at all \citep{copp2022pretrial, stevenson2022algorithmic, stevenson2018assessing, imai2023experimental}. Furthermore, pretrial risk assessments may actually drive racial inequality in the criminal legal system \citep{copp2022pretrial}, perhaps because when judges override risk tool recommendations, their decisions more often favor white defendants while punishing Black and Latinx defendants \citep{copp2022pretrial, bahl2023algorithms}. These results align with growing evidence that risk assessments are primarily beneficial to white and more affluent defendants and can have the opposite effect, intensifying harm, to Black and indigent defendants \citep{skeem2020impact, marlowe2020employing}. Although it remains relatively small, the body of research about pretrial risk assessments has grown exponentially since ProPublica's report on a particular pretrial risk assessment known as the Correctional Offender Management Profiling for Alternative Sanctions (COMPAS) \citep{angwin2016Propublica}. The report revealed that COMPAS was far more likely to erroneously predict that Black defendants would experience pretrial failure relative to white defendants and led to a number of analyses about algorithmic fairness in the criminal legal system \citep[e.g.,] [] {mitchell2021algorithmic, corbett2017algorithmic, coston2021characterizing, berk2021fairness}. 

Our goal is to investigate the accuracy of both pretrial risk assessments and judicial decisions in predicting defendant risk of reoffense or failure to appear (FTA) for trial. To accomplish this goal, we view pretrial assessments and judicial decisions as noisy observed proxies for a true outcome of interest: pretrial failure. Our proposed modeling strategy allows us to investigate the association between pretrial failure and risk factors, while simultaneously estimating misclassification rates of pretrial risk assessments and of judicial decisions as a function of defendant race. 

Our approach directly models the multistage nature of pretrial detention decisions. We incorporate model components for both the pretrial risk assessment algorithm and the judicial decision, where the judicial decision is dependent upon the algorithmic recommendation. These components comprise stages 1 and 2 of our model, respectively, allowing us to investigate decisions at multiple points in the pretrial process. Recent applied and methodological developments of multistage modeling have allowed researchers to evaluate the disparate impact of policies in the context of both pretrial decision making and police stops \citep{grossman2023racial, jung2024mitigating}. Our approach combines this concept of multistage decision-making with a growing literature that makes use of outcome misclassification models to answer questions about algorithmic fairness. Previous studies have shown that even small amounts of misclassification (4\% - 7\%) in arrest data can result in statistically significant bias against Black defendants in the COMPAS instrument \citep{fogliato2020fairness}. Similar approaches have been developed to reveal gender disparities in the context of medical diagnoses \citep{hochstedler2023statistical}. These methods have also been developed into user-friendly software packages, allowing domain experts to apply misclassification modeling techniques to novel problems \citep{combo}. These existing methods, however, rely on known misclassification rates and perfect sensitivity assumptions \citep{fogliato2020fairness} or only account for only a single observed proxy \citep{hochstedler2023statistical}. 

The stakes of this work are high. One reason for the proliferation of algorithmic pretrial risk assessment tools is the growing recognition that pretrial detention is both a major driver of the massive jail population and a hugely damaging, life altering event. More than 95\% of jail population growth over the last 40 years is attributable to the rising pretrial population \citep{zeng2018jail}, and racial/ethnic minorities and the poor are disproportionately impacted by this growth \citep{demuth2003racial, demuth2004impact, schlesinger2005racial, sutton2013structural, wooldredge2012distinguishing, wooldredge2015impact}. This is true in part because they receive higher bond amounts \citep{demuth2004impact, wooldredge2017ecological}, and can less often afford bail \citep{katz1995effect, demuth2003racial, sacks2015sentenced, schlesinger2005racial}. Jurisdictions that deploy risk assessments often do so in the hopes that they encourage judges to impose only the least restrictive means necessary to ensure public safety, functionally reducing or eliminating bond in many cases.

Once a defendant is detained pretrial, the negative impacts cascade. Not only does pretrial detention impact guilty plea decisions, case dismissals, and charge reductions \citep{hagan1975social, schlesinger2008cumulative, wooldredge2015impact}, it also increases the likelihood that defendants receive convictions and lengthens sentences for the convicted \citep{leslie2017unintended}. Pretrial detention also has many downstream effects. Even short exposure to the criminal justice system increases rates of recidivism \citep{prins2019criminogenic, marlowe2020employing}, reduces the likelihood of gaining employment, and suppresses wages among the employed \citep{dobbie2018effects}.

More research is urgently needed to understand the accuracy of pretrial risk assessments, as well as the manner in which judges interact with them. While there are many methodological proposals for evaluating algorithmic fairness, there are comparatively few existing datasets on which to test those methods. Because pretrial datasets are not widely available, many studies rely on the same one, first made available by ProPublica in 2016. Each of these studies makes important, incremental contributions to our collective understanding of the material trade offs of relying on pretrial risk assessments to solve particular policy problems. 

We expand on this growing body of research in two distinct ways: first, by introducing a novel algorithm to assess the accuracy of sequential and noisy binary decisions; second, by applying it to a countywide dataset provided by the Virginia Department of Criminal Justice Services to measure the differential misclassification of pretrial failure across racial groups. Prince William County agreed to share this data after the authors engaged in extensive qualitative research across the state. Our study is the first to make use of it. 

The remainder of this article is organized as follows. In Section 2, we introduce the motivating pretrial detention data from the Virginia Department of Criminal Justice Services. Section 3 describes the conceptual framework for our model, including both frequentist and Bayesian estimation procedures. Section 4 demonstrates the utility of our proposed methods through simulation studies across various settings.  In Section 5, we analyze the risk factors associated with pretrial failure, and simultaneously estimate the accuracy of the VPRAI and judicial decisions, with respect to defendant race. In Section 6, we provide a concluding discussion.

\section{Virginia Department of Criminal Justice Services Data} \label{data}
The study uses pretrial data from admitted persons in Prince William County, Virginia between January 2016 and December 2019 to investigate risk factors for ``pretrial failure'', defined as reoffense before trial or failure to appear (FTA) for a trial date. These records were provided by the Virginia Department of Criminal Justice Services and were collected between January 1, 2016 and December 31, 2019. We are interested in risk factors associated with relapse into criminal behavior after admittance and before trial date, a concept known as recidivism \citep{desmarais2013risk}, as well as failure to appear for trial. In this dataset, however, we do not have records of criminal activity and FTA after the initial arrest. Instead, we have two sequential and imperfect measures of this behavior. First, the risk of pretrial failure is measured using the Virginia Pretrial Risk Assessment Instrument (VPRAI). This instrument uses factors such as current charge, criminal history, employment status, and drug use to assess an individual's likelihood of reoffense or FTA. Based on the VPRAI recommendation, a judge will make a final decision to release or detain an individual before their trial. We can view the judge's decision as another imperfect measure of an individual's propensity for pretrial failure, and this outcome is based on a first-stage outcome, the VPRAI recommendation. Previous work in the criminal justice and fairness literature suggests that risk instrument recommendations and judge decisions may be frequently misclassified, and that these misclassifications may depend on the race of the defendant \citep{angwin2016Propublica}. Thus, we expect misclassification of pretrial failure to be associated with defendant race. Our overall goal is to assess the severity and direction of this misclassification, while estimating the true association between risk factors and pretrial failure.  

In our analysis, we only include individuals with charge categories for which we could reconstruct the VPRAI recommendation, including non-violent misdemeanors, driving under the influence, violent misdemeanors, and firearm charges. We also only included individuals whose race was listed as either white or Black. After excluding all records with missing values in the responses and the covariates, we had a total of 1,990 records in our dataset. Table \ref{demographics-table} displays defendant characteristics in this dataset. Of these records in our dataset, 259 individuals ($13.0\%$) received a ``detain'' VPRAI recommendation, but 1,038 defendants ($52.2\%$) were detained by the court ahead of their trial. There were 17 unique judicial officers responsible for presiding over the arraignment trials.

\begin{table}[H]
\centering
\caption{Characteristics among defendants in the Virginia Department of Criminal Justice Services dataset. Statistics are presented for the overall dataset, and with respect to defendant race, n (\%).} \label{demographics-table}
\begin{threeparttable}
\begin{tabular}{llll}
\hline
          Defendant Characteristic & Black Defendants\tnote{1} &
        White Defendants\tnote{2} &
        All\\
\hline
    \\
    Number & 987 (49.6) & 1003 (50.4) & 1990 (100)\\
    \\
    Sex (Male) & 808 (81.9) & 812 (81.0) & 1620 (81.4) \\
    Previous FTA\tnote{3}, mean (SD) & 0.16 (0.50) & 0.07 (0.36) & 0.12 (0.43)\\
    Employment status (unemployed) & 349 (35.3) & 287 (28.6) & 636 (32.0) \\
    History of drug abuse & 439 (44.4) & 340 (34.4) & 779 (77.7) \\
    Previous violent arrests\tnote{4}, mean (SD) & 1.44 (2.20) & 0.75 (1.60) & 1.09 (1.95)\\
    \\
    VPRAI detain recommendation & 189 (19.1) & 70 (7.0) & 259 (13.0) \\
    Court detention decision & 587 (59.5) & 451 (45.0) & 1038 (52.2)
              \\
\hline  
\end{tabular}
\begin{tablenotes}
\item[1] For all characteristics aside from "Number", the percentage in this column is out of the total number of Black defendants. 
\item[2] For all characteristics aside from "Number", the percentage in this column is out of the total number of white defendants. 
\item[3] Number of previous failures to appear (FTA) for trial dates.
\item[4] Number of previous violent arrests.
\end{tablenotes}
\end{threeparttable}
\end{table}

\section{A Two-Stage Misclassification Model} \label{multistage-model}
\begin{figure}[H]
\begin{center}
\includegraphics[scale=0.7]{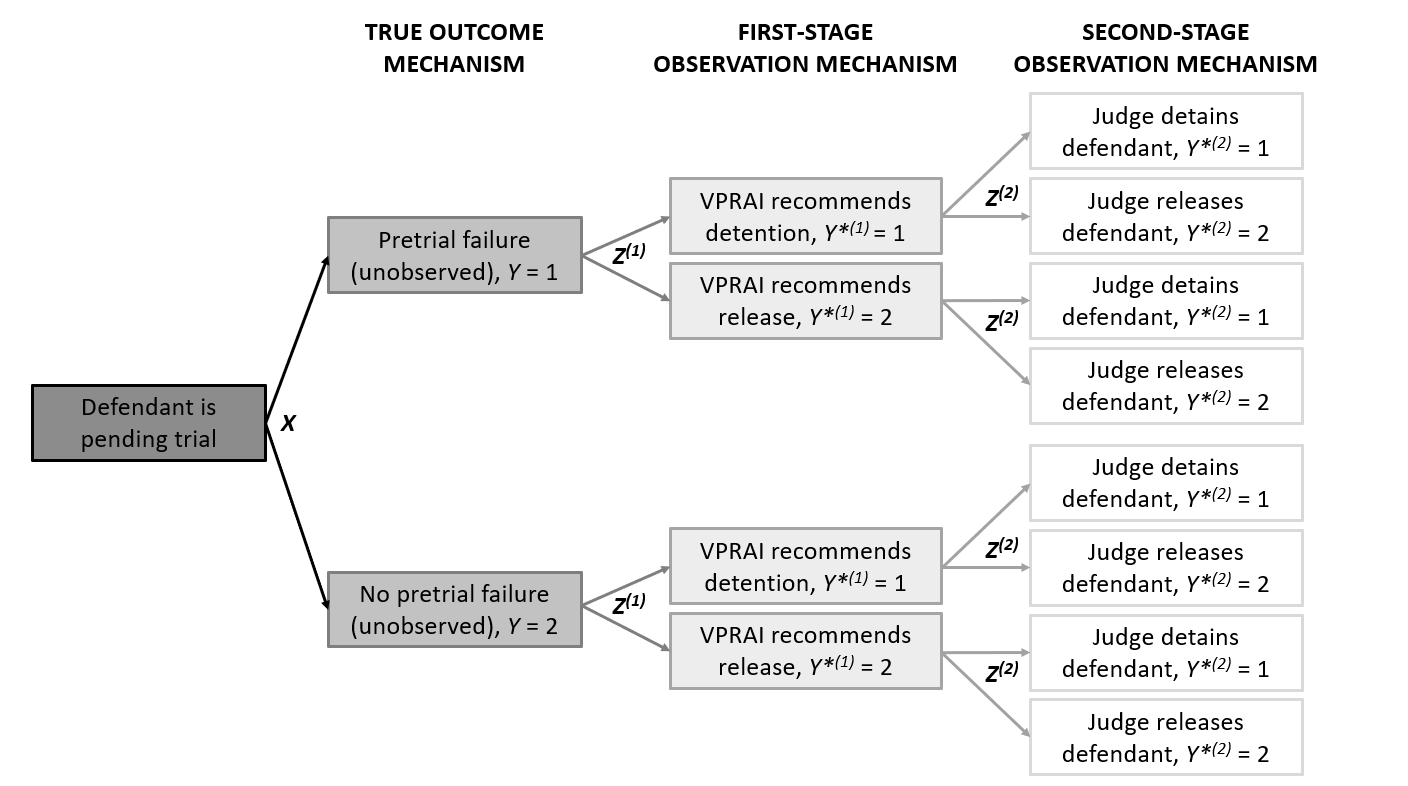}

\caption{Diagram of the assumed data structure for a two-stage misclassification model. Here, $\boldsymbol{X}$ is a set of predictors related to pretrial failure. $Y$ represents true pretrial failure status. $Y$ is a latent variable, meaning that it is not possible to observe this variable directly. Instead, we rely on imperfect proxies for pretrial failure. $\boldsymbol{Z^{(1)}}$ is a set of predictors related to the VPRAI recommendation, conditional on true (unobserved) pretrial failure status. $Y^{*(1)}$ represents the VPRAI recommendation. Similarly, $\boldsymbol{Z^{(2)}}$ is a set of predictors related to the judge's decision, given the VPRAI recommendation and true (unobserved) pretrial failure status. $Y^{*(2)}$ represents the judge's decision. }\label{multistage_conceptual_framework_figure}
\end{center}
\end{figure}

The misclassification model first introduced in \cite{hochstedler2023statistical} can be extended to a multistage framework, in the context of pretrial risk assessment. Let $Y = j$ denote an observation's true pretrial failure status, taking values $j \in \{1, 2\}$. Here $Y = 1$ indicates that an individual would truly experience a pretrial failure and $Y = 2$ indicates that an individual would \textit{not} experience a pretrial failure. We are interested in the relationship between $Y$ and predictors $\boldsymbol{X}$ related to pretrial failure. This relationship is called the \textit{true outcome mechanism}. Instead observing pretrial failure directly, we have two sequential and imperfect measurements of $Y$ -- the VPRAI recommendation and the judge's decision. Let $Y^{*(1)}$ denote the VPRAI algorithm recommendation, which is observed in the first stage of the data generating process, taking values $k^{(1)} \in \{ 1, 2 \}$. Here $Y^{*(1)} = 1$ denotes a VPRAI "detain" recommendation and $Y^{*(1)} = 2$ denotes a VPRAI "release" recommendation. $Y^{*(2)}$ denotes the judge's pretrial decision, which is observed in the second stage of the data generating process. $Y^{*(2)}$ takes values $k^{(2)} \in \{ 1, 2 \}$ where $Y^{*(2)} = 1$ indicates that the judge detained the individual and $Y^{*(2)} = 2$ indicates that the judge released the individual. Because $Y^{*(1)}$ and $Y^{*(2)}$ do not always match an individual's true pretrial failure status, $Y$, we consider them to be noisy indicators of the outcome of interest. Let $\boldsymbol{Z^{(1)}}$ denote a matrix of predictors related to the misclassification of $Y^{*(1)}$. Similarly, $\boldsymbol{Z^{(2)}}$ denote a matrix of predictors related to the misclassification of $Y^{*(2)}$. The mechanism that generates the first-stage observed outcome, $Y^{*(1)}$, given the true outcome, $Y$, is called the \textit{first-stage observation mechanism}. The mechanism that generates the second-stage observed outcome $Y^{*(2)}$, conditional on the first-stage observed outcome, $Y^{*(1)}$, and the true outcome, $Y$, is called the \textit{first-stage observation mechanism}. Figure \ref{multistage_conceptual_framework_figure} displays the conceptual model for a two-stage misclassification model. 

The conceptual process illustrated in Figure \ref{multistage_conceptual_framework_figure} can be expressed mathematically as
\begin{equation}
\begin{aligned}
\label{eq:multistage_conceptual_framework_eq-1}
\text{True outcome mechanism: } &\; \text{logit}\{ P(Y = 1 | \boldsymbol{X} ; \boldsymbol{\beta}) \} = \beta_{0} + \boldsymbol{\beta}_{X} \boldsymbol{X}.
\end{aligned}
\end{equation}

\begin{equation}
\begin{aligned}
\label{eq:multistage_conceptual_framework_eq-2}
\text{ First-Stage observation mechanisms: } &\; \text{logit}\{ P(Y^{*(1)} = 1 | Y = 1, \boldsymbol{Z^{(1)}} ; \boldsymbol{\gamma}^{(1)}) \} = \gamma^{(1)}_{110} + \boldsymbol{\gamma^{(1)}_{11Z^{(1)}}} \boldsymbol{Z^{(1)}}\\
&\; \text{logit}\{ P(Y^{*(1)} = 1 | Y = 2, \boldsymbol{Z^{(1)}} ; \boldsymbol{\gamma}^{(1)}) \} = \gamma^{(1)}_{120} + \boldsymbol{\gamma^{(1)}_{12Z^{(1)}}} \boldsymbol{Z^{(1)}}.
\end{aligned}
\end{equation}

\begin{equation}
\begin{aligned}
\label{eq:multistage_conceptual_framework_eq-3}
\text{Second-Stage observation mechanisms: } &\; \text{logit}\{ P(Y^{*(2)} = 1 | Y^{*(1)} = 1, Y = 1, \boldsymbol{Z^{(2)}} ; \boldsymbol{\gamma}^{(2)}) \}  \\ &\; \hspace{1.5em} = \gamma^{(1)}_{1110} + \boldsymbol{\gamma^{(2)}_{111Z^{(2)}}} \boldsymbol{Z^{(2)}}\\
&\; \text{logit}\{ P(Y^{*(2)} = 1 | Y^{*(1)} = 2, Y = 1, \boldsymbol{Z^{(2)}} ; \boldsymbol{\gamma}^{(2)}) \} \\ &\; \hspace{1.5em} = \gamma^{(1)}_{1210} + \boldsymbol{\gamma^{(2)}_{121Z^{(2)}}} \boldsymbol{Z^{(2)}}\\
&\; \text{logit}\{ P(Y^{*(2)} = 1 | Y^{*(1)} = 1, Y = 2, \boldsymbol{Z^{(2)}} ; \boldsymbol{\gamma}^{(2)}) \} \\ &\; \hspace{1.5em} = \gamma^{(1)}_{1120} + \boldsymbol{\gamma^{(2)}_{112Z^{(2)}}} \boldsymbol{Z^{(2)}}\\
&\; \text{logit}\{ P(Y^{*(2)} = 1 | Y^{*(1)} = 2, Y = 2, \boldsymbol{Z^{(2)}} ; \boldsymbol{\gamma}^{(2)}) \} \\ &\; \hspace{1.5em} = \gamma^{(1)}_{1220} + \boldsymbol{\gamma^{(2)}_{122Z^{(2)}}} \boldsymbol{Z^{(2)}}.
\end{aligned}
\end{equation}
In this setup, category $2$ is the reference category for the true outcome mechanism and all corresponding $\boldsymbol{\beta}$ parameters are set to 0. Similarly, in the observation mechanisms, the reference category is $2$ so all corresponding $\boldsymbol{\gamma}$ parameters are set to 0. The main parameters of interest are the $\boldsymbol{\beta}$ terms, because these parameters describe the relationship between the risk factors of interest and the probability of pretrial failure. The $\boldsymbol{\gamma}^{(1)}$ parameters describe how $\boldsymbol{Z^{(1)}}$ relates to the probability of a VPRAI detention recommendation, given true pretrial status. The $\boldsymbol{\gamma}^{(2)}$ parameters describe how $\boldsymbol{Z^{(2)}}$ relates to the probability of a judge deciding to detain an individual, given the VPRAI detention recommendation and true pretrial status. Based on these interpretations, the $\boldsymbol{\gamma}$ parameters reveal how misclassification rates in the VPRAI recommendations and judge decisions may vary based on factors in $\boldsymbol{Z^{(1)}}$ and $\boldsymbol{Z^{(2)}}$.

If potential misclassification in VPRAI recommendations and/or judge decisions is ignored, and either of these observed outcomes is used as a naive proxy for pretrial failure, we expect bias in association parameter estimates. In particular if a model of $P(Y^{*(1)} = 1 | \boldsymbol{X})$ or $P(Y^{*(2)} = 1 | \boldsymbol{X})$ is fit, rather than a model of $P(Y = 1 | \boldsymbol{X})$, we expect naive estimators that are not equal to the true association parameters of interest, $\boldsymbol{\beta}$. This bias can be especially severe if misclassification in the observed outcomes is covariate-dependent \citep{beesley2020statistical}.

The probability of true pretrial failure status category $j$ for individual $i$ is by denoted $\pi_{ij}$. The probability of VPRAI recommendation $k$, conditional on true pretrial failure status $j$ is denoted $\pi^{*(1)}_{i k j}$. The probability of judge decision $\ell$, conditional on VPRAI recommendation $k$ and true pretrial failure status $j$ is denoted $\pi^{*(2)}_{i \ell k j}$. Using (\ref{eq:multistage_conceptual_framework_eq-1}), (\ref{eq:multistage_conceptual_framework_eq-2}), and (\ref{eq:multistage_conceptual_framework_eq-3}), we can express these response probabilities as follows:  
\begin{flalign}
\begin{aligned}
\label{eq:multistage_response_probabilities_eq}
P(Y_i = j | \boldsymbol{X_i} ; \boldsymbol{\beta}) = &\; \; \pi_{ij} = \frac{\text{exp}\{\beta_{j0} + \boldsymbol{\beta_{jX} X_i}\}}{1 + \text{exp}\{\beta_{j0} + \boldsymbol{\beta_{jX} X_i}\}} \\
P(Y^{*(1)}_i = k | Y_i = j, \boldsymbol{Z^{(1)}} ; \boldsymbol{\gamma^{(1)}}) = &\; \pi^{*(1)}_{i k j} = \frac{\text{exp}\{\gamma^{(1)}_{kj0} + \boldsymbol{\gamma^{(1)}_{kjZ} Z^{(1)}_i}\}}{1 + \text{exp}\{\gamma^{(1)}_{kj0} + \boldsymbol{\gamma^{(1)}_{kjZ} Z^{(1)}_i}\}} \\
P(Y^{*(2)}_i = \ell | Y^{*(1)}_i = k, Y_i = j, \boldsymbol{Z^{(2)}} ; \boldsymbol{\gamma^{(2)}}) = &\; \pi^{*(2)}_{i \ell k j} = \frac{\text{exp}\{\gamma^{(2)}_{\ell kj0} + \boldsymbol{\gamma^{(2)}_{\ell kjZ} Z^{(2)}_i}\}}{1 + \text{exp}\{\gamma^{(2)}_{\ell kj0} + \boldsymbol{\gamma^{(2)}_{\ell kjZ} Z^{(2)}_i}\}}.
\end{aligned}
\end{flalign}

These quantities can be computed for all $N$ individuals in the sample, with each individual denoted as $i \in 1, \dots, N$.

When $k$ and $j$ are both equal to the reference category, $\frac{1}{N}\sum_{i = 1}^N \pi^{*(1)}_{i22} = \pi^{*(1)}_{22}$ measures the average specificity of the VPRAI algorithm. When $k$ and $j$ are both equal to the 1, $\frac{1}{N}\sum_{i = 1}^N \pi^{*(1)}_{i11} = \pi^{*(1)}_{11}$ measures the average sensitivity of the VPRAI algorithm. Thus, (\ref{eq:multistage_response_probabilities_eq}) allows us to model the sensitivity and specificity of the VPRAI recommendation algorithm based on a set of covariates, $ Z^{(1)}$.

By marginalizing over the VPRAI recommendation outcome, we can also compute the average sensitivity and specificity of judge decisions based on the set of covariates, $ Z^{(2)}$. When $\ell$ and $j$ are both equal to $2$, $\frac{1}{N}\sum_{i = 1}^N \sum_{k = 1}^2 \pi^{*(2)}_{i 2 k 2} \pi^{*(1)}_{ik2} = \pi^{*(2)}_{i22}$ measures the average specificity of the judge's decisions. Similarly, when $\ell$ and $j$ are both equal to $1$, $\frac{1}{N}\sum_{i = 1}^N \sum_{k = 1}^2 \pi^{*(2)}_{i 1 k 1} \pi^{*(1)}_{ik1} = \pi^{*(2)}_{i11}$ provides the average sensitivity of the judge's decisions.

This model has been described for two outcome stages, but it can be extended to an arbitrary number of outcome stages. This extension is available in Appendix \ref{appendix-multistage-model}.

\subsection{Estimation methods}
Our objective is to estimate parameters $\boldsymbol{\beta}$ and $\boldsymbol{\gamma}$ from the proposed model in (\ref{eq:multistage_conceptual_framework_eq-1}), (\ref{eq:multistage_conceptual_framework_eq-2}), and (\ref{eq:multistage_conceptual_framework_eq-3}). We developed two estimation methods for the two-stage misclassification model. First, we present an Expectation-Maximization algorithm to jointly estimate $\boldsymbol{\beta}$ and $\boldsymbol{\gamma}$ \citep{dempster1977maximum}. Details on this algorithm are provided in Appendix \ref{estimation-details-em}. Our second estimation method is a Markov Chain Monte Carlo (MCMC) procedure. Details on this Bayesian estimation method are available in Appendix \ref{estimation-details-mcmc}. For both of the proposed methods, we must account for the known issue of ``label switching'', which causes our problem to have two plausible solutions, rather than one \citep{betancourt2017identifying, stephens2000dealing}. Our label switching correction relies on the assumption that the sum of the first-stage sensitivity and specificity estimates is at least 1. This assumption is realistic in practice, because it requires the VPRAI recommendation algorithm to perform better than chance, on average. More information on label switching and our proposed correction strategy is provided in Appendix \ref{label-switching}. The performance of our proposed estimation techniques was evaluated through simulation studies, the results of which are available in Appendix \ref{simulations}.  

Both the EM algorithm and MCMC estimation strategies are available in the open-source R Package \textit{COMBO} \citep{combo}. This R Package also includes the label switching correction. 

\section{Evaluating the Accuracy of Pretrial Assessments and Judicial Decisions} \label{example}
In this section, we apply our model to investigate risk factors for reoffense or FTA in the presence of two sequential and dependent noisy outcomes: VPRAI recommendations and judge decisions. 

Our first stage response variable, $Y^{*(1)}$, is the VPRAI recommendation, dichotomized into ``detain'' and ``release'' recommendation categories. Our second-stage response variable, $Y^{*(2)}$, is the court's final decision to detain or release a defendant before their trial. We assessed the association of pretrial failure with risk factors, $\boldsymbol{X}$, including the number of prior failures to appear (FTA) for trial, employment status, drug abuse history, and the number of previous violent arrests. Employment and drug history variables were coded as binary indicators. In addition, we assessed the association of defendant race, $\boldsymbol{Z^{(1)}}$ and $\boldsymbol{Z^{(2)}}$, to each of the observed outcomes, conditional on earlier stage outcomes and on true pretrial failure status.

We estimated model parameters using the EM algorithm and label switching correction, defined in Section \ref{multistage-model}. Parameter estimates and standard errors are provided in Table \ref{applied-results-table-beta}, Table \ref{applied-results-table-gamma1}, and Table \ref{applied-results-table-gamma2}. With the exception of the intercept and drug history association parameters, the $\boldsymbol{\beta}$ estimates in the misclassification model are all attenuated compared to the naive model. We find that increased numbers of previous FTA, unemployment, a history of drug abuse, and increased numbers of violent arrests are all associated with true incidence of recidivism or FTA in this data set. For the first-stage outcome, we find that, given true recidivism or FTA, Black defendants are more likely to have a VPRAI detention recommendation than white defendants. Moreover, given no reoffense or FTA, Black defendants are \textit{still} more likely to have a VPRAI detention recommendation than white defendants. This trend remains for court decisions. Given no pretrial failure and regardless of VPRAI recommendations, Black individuals are more likely to be detained by the court than white individuals. Similarly, given pretrial failure and a VPRAI detention recommendation, Black defendants are more likely to be detained before their trial than white defendants. Black defendants are also more likely to be detained before their trial than white defendants in the case of true recidivism or FTA and a VPRAI recommendation of ``release'', though this parameter estimate has extremely high standard error due to perfect separation, or perfect prediction of the outcome by race \citep{mansournia2018separation}. In fact, conditional on true recidivism or FTA and a VPRAI ``release'' recommendation, our model estimates that $100\%$ of Black defendants are detained by judges.  

\begin{table}[H]
\centering
\caption{Parameter estimates and standard errors for parameters in the \textit{true outcome mechanism} from the study using the Virginia Department of Criminal Justice Services dataset.``EM'' estimates were computed using the \textit{COMBO} R Package \citep{combo}. The ``Naive Analysis'' results were obtained by running a two-stage model that does not account for outcome misclassification. Estimates marked with a ``-'' are not obtained by the given estimation method.} \label{applied-results-table-beta}
\begin{threeparttable}
\begin{tabular}{llrrrrrrrrrrrr}
\hline
           && \multicolumn{2}{c}{EM}       &
        \multicolumn{2}{c}{Naive Analysis}\\
        \cline{3-6}
 $\boldsymbol{X}$ Coefficients && \multicolumn{1}{c}{Est.} & \multicolumn{1}{c}{SE} & \multicolumn{1}{c}{Est.} & \multicolumn{1}{c}{SE}\\
\hline
    \\
    Intercept && -3.512 & 0.105 & -3.744 & 0.170 \\
    Number of previous FTAs\tnote{1}  && 1.224 & 0.218 & 1.026 & 0.134 \\
    Employment status \tnote{2} && 0.732 & 0.056 & 0.676 & 0.152 \\
    History of drug use \tnote{3} && 1.968 & 0.126 & 1.743 & 0.167 \\
    Number of previous violent arrests \tnote{4} && 0.280 & 0.022 & 0.256 & 0.030 \\
              \\
\hline  
\end{tabular}
\begin{tablenotes}
\item[1] $\beta_{FTA}$ refers to the association between a defendant's number of previous failures to appear (FTA) and risk of pretrial failure.
\item[2] $\beta_{unemployed}$ refers to the association between employment status (reference =  employed or unable to work) and risk of pretrial failure.
\item[3] $\beta_{drug}$ refers to the association between drug abuse history (reference = no history of drug abuse) and risk of pretrial failure.
\item[4] $\beta_{violent}$ refers to the association between a defendant's number of previous violent arrests and risk of pretrial failure.
\end{tablenotes}
\end{threeparttable}
\end{table}

\begin{table}[H]
\centering
\caption{Parameter estimates and standard errors for parameters in the \textit{first-stage observed outcome mechanism} from the study using the Virginia  Department of Criminal Justice Services dataset.``EM'' estimates were computed using the \textit{COMBO} R Package \citep{combo}. The ``Naive Analysis'' results were obtained by running a two-stage model that does not account for outcome misclassification. Estimates marked with a ``-'' are not obtained by the given estimation method.} \label{applied-results-table-gamma1}
\begin{threeparttable}
\begin{tabular}{clrrrrrrrrrrrr}
\hline
           && \multicolumn{2}{c}{EM}       &
        \multicolumn{2}{c}{Naive Analysis}\\
        \cline{3-6}
 $\boldsymbol{Z^{(1)}}$ Coefficients && \multicolumn{1}{c}{Est.} & \multicolumn{1}{c}{SE} & \multicolumn{1}{c}{Est.} & \multicolumn{1}{c}{SE}\\
\hline
    \\
    Intercept$^{(1)}_{11}$ \tnote{1} && -0.029 & 0.199  & - & -\\
    \\
    Race$^{(1)}_{11}$ \tnote{2} && 1.843 & 0.376 & - & -\\
    \\
    Intercept$^{(1)}_{12}$ \tnote{3} && -20.270 & 0.605 & - & - \\
    \\
    Race$^{(1)}_{12}$ \tnote{4} && 15.341 & 0.603 & - & - \\
              \\
\hline  
\end{tabular}
\begin{tablenotes}
\item[1] $\gamma^{(1)}_{110}$ is the intercept term in the first-stage observation mechanism, conditional on $Y = 1$.
\item[2] $\gamma^{(1)}_{11,race}$ refers to the association between a defendant's race (reference = white) and the probability of a VPRAI "detain'' recommendation, given that the individual would truly experience pretrial failure, $Y = 1$.
\item[3] $\gamma^{(1)}_{120}$ is the intercept term in the first-stage observation mechanism, conditional on $Y = 2$.
\item[4] $\gamma^{(1)}_{12,race}$ refers to the association between a defendant's race (reference = white) and the probability of a VPRAI "detain'' recommendation, given that the individual would \textit{not} truly experience pretrial failure, $Y = 2$.
\end{tablenotes}
\end{threeparttable}
\end{table}

\begin{table}[H]
\centering
\caption{Parameter estimates and standard errors for parameters in the \textit{second-stage observed outcome mechanism} from the study using the Virginia  Department of Criminal Justice Services dataset.``EM'' estimates were computed using the \textit{COMBO} R Package \citep{combo}. The ``Naive Analysis'' results were obtained by running a two-stage model that does not account for outcome misclassification. Estimates marked with a ``-'' are not obtained by the given estimation method.} \label{applied-results-table-gamma2}
\begin{threeparttable}
\begin{tabular}{clrrrrrrrrrrrr}
\hline
           &&&& \multicolumn{2}{c}{EM}       & &
        \multicolumn{2}{c}{Naive Analysis}\\
        \cline{3-9}
$\boldsymbol{Z^{(1)}}$ Coefficients &&&& \multicolumn{1}{c}{Est.} & \multicolumn{1}{c}{SE} & & \multicolumn{1}{c}{Est.} & \multicolumn{1}{c}{SE}\\
\hline
\\
    Intercept$^{(2)}_{111}$ \tnote{1} &&&& 1.576 & 0.298 && 1.576 & 0.317 \\[.3cm]
    
    Race$^{(2)}_{111}$ \tnote{2} &&&& 0.327 & 0.374 && 0.352 & 0.385\\[.3cm]
    
    Intercept$^{(2)}_{121}$ \tnote{3} &&&& 0.892 & 0.203 && - & - \\[.3cm]
    
   Race$^{(2)}_{121}$ \tnote{4}&&&& 15.645 & 182.618 && - & - \\[.3cm]

    Intercept$^{(2)}_{112}$ \tnote{5} &&&& -6.492 & 1.858 && - & -\\[.3cm]
    
    Race$^{(2)}_{112}$ \tnote{6}&&&& 9.822 & 1.859 && - & -\\[.3cm]
    
    Intercept$^{(2)}_{122}$ \tnote{7} &&&& -0.418 & 0.018 & & -0.318 & 0.066 \\[.3cm]
    
    Race$^{(2)}_{122}$ \tnote{8}&&&& 0.459 & 0.018 && 0.433 & 0.097 \\[.3cm]
\hline  
\end{tabular}
\begin{tablenotes}
\item[1] $\gamma^{(2)}_{1110}$ is the intercept term in the second-stage observation mechanism, conditional on $Y^{(1)}= 1$ and $Y = 1$.
\item[2] $\gamma^{(2)}_{111,race}$ refers to the association between a defendant's race (reference = white) and the probability of a judge "detention'' decision, given that the VPRAI recommended detention and that the individual would truly experience pretrial failure, $Y = 1$.
\item[3] $\gamma^{(2)}_{1210}$ is the intercept term in the second-stage observation mechanism, conditional on $Y^{(1)}= 2$ and $Y = 1$.
\item[4] $\gamma^{(2)}_{121,race}$ refers to the association between a defendant's race (reference = white) and the probability of a judge "detention'' decision, given that the VPRAI recommended release and that the individual would truly experience pretrial failure, $Y = 1$.
\item[5] $\gamma^{(2)}_{1120}$ is the intercept term in the second-stage observation mechanism, conditional on $Y^{(1)}= 1$ and $Y = 2$.
\item[6] $\gamma^{(2)}_{112,race}$ refers to the association between a defendant's race (reference = white) and the probability of a judge "detention'' decision, given that the VPRAI recommended detention and that the individual would \textit{not} truly experience pretrial failure, $Y = 2$.
\item[7] $\gamma^{(2)}_{1220}$ is the intercept term in the second-stage observation mechanism, conditional on $Y^{(1)}= 2$ and $Y = 2$.
\item[8] $\gamma^{(2)}_{112,race}$ refers to the association between a defendant's race (reference = white) and the probability of a judge "detention'' decision, given that the VPRAI recommended release and that the individual would \textit{not} truly experience pretrial failure, $Y = 2$.
\end{tablenotes}
\end{threeparttable}
\end{table}

We also use the EM Algorithm parameter estimates to assess VPRAI sensitivity and specificity, as well as the fairness and accuracy of judge decisions. In the sample, we estimate a pretrial failure rate of $17.9\%$. The VPRAI appears to have moderate sensitivity and near-perfect specificity; we estimate that the VPRAI correctly recommends detention for $67.5\%$ of defendants and correctly recommends release for $99.6\%$ of defendants. This sensitivity rate, however, differs by defendant race. Among Black defendants who are expected to reoffend or fail to appear for their trial date, the VPRAI recommends pretrial detention $86.0\%$ of the time. Among white defendants, this rate drops to just $49.3\%$. Moving to judge decisions, we estimate that, among individuals who had a VPRAI detention recommendation, judges correctly detain defendants in $84.9\%$ of cases. However, among defendants who received a VPRAI release recommendation, we estimate that judges correctly release defendants just $54.7\%$ of the time. Again, these rates differ by defendant race. Among defendants estimated to truly reoffend or not appear for trial and who were given a VPRAI detention recommendation, a court detention decision is received by $82.9\%$ of white defendants and $87.0\%$ of Black defendants. Among white defendants who are not expected to reoffend or fail to appear and who have VPRAI ``release'' recommendations, we estimate that $60.3\%$ are, in fact, released before their trial. Among Black defendants, this proportion drops to just $49.0\%$. Collapsing across VPRAI recommendations, we estimate that judges appropriately detain white individuals in $76.8\%$ of cases and appropriately detain Black individuals in $88.8\%$ of cases, suggesting that white defendants may be given ``the benefit of the doubt'' more often than Black defendants. Similarly, we estimate that white defendants are wrongfully detained by the court in $39.7\%$ of cases, but wrongful detentions happen in as many as $51.4\%$ of cases involving Black defendants.

\section{Discussion}
In this work, we assess the accuracy of pretrial risk assessment recommendations and judicial decisions in predicting defendant pretrial failure and find that Black defendants are more often misclassified—both by the VPRAI and, to a greater extent, judicial officers—relative to their white counterparts. Practically, this means that, conditioned on the likelihood of pretrial failure, the VPRAI more often recommends detention for Black defendants, and that judges are more likely to detain Black defendants, regardless of VPRAI recommendation or likelihood of pretrial failure. This suggests that pretrial risk assessments may actually institutionalize or exacerbate the very racial inequality they are intended to combat. These results have implications not only for Virginia, but also for every jurisdiction that administers pretrial risk assessments.

Beyond this specific application, our algorithm has a wide array of use cases in other high-stakes policy settings. Many public sector algorithms estimate binary outcomes to aid human decision-making, including child welfare intervention systems \citep{chouldechova2018case}, welfare fraud detection algorithms \citep{eubanks2018automating}, and public housing allocation systems \citep{balagot2019homeless, schneider2020locked}. The decisions these algorithms and agencies make can change the course of a person’s life. Our algorithm offers a method of exposing instances of misclassification that may systematically disadvantage certain groups of people. While similar approaches of measuring such disparities exist, they require known misclassification rates and perfect sensitivity \citep{fogliato2020fairness}, or only account for only a single observed proxy \citep{hochstedler2023statistical}. Our methods have the additional strengths of not requiring gold standard labels and of being a multi-stage generalization of the work of \cite{hochstedler2023statistical}. Specifically, we can handle data generation processes that are comprised of multiple dependent and sequential misclassified binary outcomes. 

Further generalizations of our methods are still possible in future work. For example, our work assumes that predictor variables are correctly measured, which is unlikely in practice. A more complex modeling structure would be required to account for imperfect arrest records or demographic variables. In addition, our method is limited to sequential and dependent binary outcomes. If categorical or continuous outcomes are present in a research design, further extensions of this work would be required. Finally, our methods make the implicit assumption that all court decisions are independent of one another. It is perhaps more realistic to assume that decisions made by a given judicial officer are more correlated with one another than with the decisions made by a different judicial officer. Dependency structures among decision-makers have not yet been incorporated into our models.

\subsection*{Acknowledgements}
The authors would like to thank Karen Levy and Solon Barocas for their insights into pretrial detention processes. Data used in the two-stage misclassification model analysis was provided by the Virginia Department of Criminal Justice Services. 

\subsection*{Data Availability Statement}
This paper uses data from a Virginia state agency. The data use agreement (DUA) does not permit individuals who are not on the DUA to access the data. We do, however, provide computer code to generate data similar to that used in our simulation studies. A demonstration of this code and analysis is available on the first author's public GitHub profile. 

\subsection*{Funding Statement}
Funding support for Kimberly A. H. Webb was provided by the LinkedIn and Cornell Ann S. Bowers College of Computing and Information Science strategic partnership PhD Award. Funding support for Sarah A. Riley was provided by the John D. and Catherine T. MacArthur Foundation. Martin T. Wells was supported by NIH awards U19AI111143-07 and P01-AI159402.

\subsection*{Conflict of Interest Disclosure}
The authors report there are no competing interests to declare.

\subsection*{Ethics Approval Statement}
This research was approved by Cornell University's Institutional Review Board.

\newpage

\begin{appendices}

\section{A Multistage Misclassification Model}\label{appendix-multistage-model}
The misclassification model first introduced in \cite{hochstedler2023statistical} can be extended to a multistage framework. Let $Y = j$ denote an observation's true outcome status, taking values $j \in \{1, 2\}$ and we are interested in the relationship between $Y$ and a matrix of predictors $\boldsymbol{X}$. Instead of obtaining just one potentially misclassified measurement of $Y$ as in \cite{hochstedler2023statistical}, we now have $a$ sequential imperfect measurements of $Y$. Let $Y^{*(a)}$ denote the observed outcome from stage $a$ of the data generating process, taking values $k^{(a)} \in \{ 1, 2 \}$. Let $\boldsymbol{Z^{(a)}}$ denote a matrix of predictors related to the misclassification of $Y^{*(a)}$. The mechanism that generates the observed outcome, $Y^{*(a)}$, given the true outcome, $Y$, and all earlier-stage observed outcomes, $Y^{*(a - 1)} \dots Y^{*(1)}$, is called the $a^{th}$\textit{-stage observation mechanism}.

The conceptual process can be expressed mathematically as
\begin{equation}
\begin{aligned}
\label{eq:appendix-multistage_conceptual_framework_eq}
\text{True outcome mechanism: } &\; \text{logit}\{ P(Y = j | \boldsymbol{X} ; \boldsymbol{\beta}) \} = \beta_{j0} + \boldsymbol{\beta}_{jX} \boldsymbol{X} \\
a^{th} \text{ Observation mechanisms: } &\; \text{logit}\{ P(Y^{*(a)} = k^{(a)} | \mathcal{Y}^{*(a - 1)}, Y = 1, \boldsymbol{Z^{(a)}} ; \boldsymbol{\gamma}) \}  
\\ &\; = \gamma^{(a)}_{\mathscr{k}^{(a)} 10} + \boldsymbol{\gamma^{(a)}_{\mathscr{k}^{(a)}1Z^{(a)}}} \boldsymbol{Z^{(a)}}\\
&\; \text{logit}\{ P(Y^{*(a)} = k^{(a)} | \mathcal{Y}^{*(a - 1)}, Y = 2, \boldsymbol{Z^{(a)}} ; \boldsymbol{\gamma}) \}  
\\ &\; = \gamma^{(a)}_{\mathscr{k}^{(a)} 20} + \boldsymbol{\gamma^{(a)}_{\mathscr{k}^{(a)}2Z^{(a)}}} \boldsymbol{Z^{(a)}}
\end{aligned}
\end{equation}
where $\mathscr{k}^{(a)} = \{ k^{(a)}, k^{(a - 1)}, \dots, k^{(1)} \}$ and $\mathcal{Y}^{*(a - 1)} = \{ Y^{*(a - 1)} = k^{(a - 1)}, \dots, Y^{*(1)} = k^{(1)} \}$. In this setup, category $2$ is the reference category for the true outcome mechanism and all corresponding $\boldsymbol{\beta}$ parameters are set to 0. Similarly, in the observation mechanisms, the reference category is $Y^{*(a)} = 2$ so all corresponding $\boldsymbol{\gamma}$ parameters are set to 0. The probability of true outcome category $j$ for individual $i$ is by denoted $\pi_{ij}$. The probability of $a^{th}$ stage observed outcome category $k$, conditional on all earlier-stage outcomes $\mathscr{k}^{(a)}$ and the true outcome $j$ is denoted $\pi^{*(a)}_{i \mathscr{k}^{(a)} j}$. Using (\ref{eq:appendix-multistage_conceptual_framework_eq}), we can express both of these response probabilities as follows:  
\begin{flalign}
\begin{aligned}
\label{eq:appendix-multistage_response_probabilities_eq}
P(Y^{*(a)} = k^{(a)} | \boldsymbol{X}, \boldsymbol{Z^{(a)}}, \dots, \boldsymbol{Z^{(1})})   &\;
= \sum_{j = 1}^2 \sum_{k^{(a-1)}, \dots, k^{(1)} = 1}^2 \Bigl( P(Y^{*(a)} = k^{(a)} | \mathcal{Y}^{*(a - 1)}, Y = j, \boldsymbol{Z^{(a)}}, \dots, \boldsymbol{Z^{(1)}}; \boldsymbol{\gamma}) \\
&\qquad\phantom{a} \times P(Y^{*(a - 1)} = k^{(a - 1)} | \mathcal{Y}^{*(a - 2)}, Y = j, \boldsymbol{Z^{(a - 1)}}, \dots, \boldsymbol{Z^{(1)}}; \boldsymbol{\gamma}) \times \dots \\
&\qquad\phantom{a} \times P(Y^{*(1)} = k^{(1)} | Y = j, \boldsymbol{Z^{(1)}}; \boldsymbol{\gamma}) \times P(Y = j | \boldsymbol{X} ; \boldsymbol{\beta}) \Bigr)
\\ & = \sum_{j = 1}^2 \sum_{k^{(a-1)}, \dots, k^{(1)} = 1}^2 \Bigl( \pi^{*(a)}_{\mathscr{k}^{(a)}j} \times \pi^{*(a - 1)}_{\mathscr{k}^{(a - 1)}j} \times \dots \times \pi^{*(1)}_{\mathscr{k}^{(1)}j} \times \pi_{j} \Bigr).
\end{aligned}
\end{flalign}

We define the probability of observing outcomes $\mathscr{k}^{(a)}$ in stage $a$ of the model using the model structure
\begin{equation}
\begin{aligned}
\label{eq:multistage_p_obs_Ystar}
P(Y^{*(a)} = k^{(a)} | \boldsymbol{X}, \boldsymbol{Z^{(a)}}, \dots, \boldsymbol{Z^{(1})})   &\;
= \sum_{j = 1}^2 \sum_{k^{(a-1)}, \dots, k^{(1)} = 1}^2 \Bigl( P(Y^{*(a)} = k^{(a)} | \mathcal{Y}^{*(a - 1)}, Y = j, \boldsymbol{Z^{(a)}}, \dots, \boldsymbol{Z^{(1)}}; \boldsymbol{\gamma}) \\
&\qquad\phantom{a} \times P(Y^{*(a - 1)} = k^{(a - 1)} | \mathcal{Y}^{*(a - 2)}, Y = j, \boldsymbol{Z^{(a - 1)}}, \dots, \boldsymbol{Z^{(1)}}; \boldsymbol{\gamma}) \times \dots \\
&\qquad\phantom{a} \times P(Y^{*(1)} = k^{(1)} | Y = j, \boldsymbol{Z^{(1)}}; \boldsymbol{\gamma}) \times P(Y = j | \boldsymbol{X} ; \boldsymbol{\beta}) \Bigr)
\\ & = \sum_{j = 1}^2 \sum_{k^{(a-1)}, \dots, k^{(1)} = 1}^2 \Bigl( \pi^{*(a)}_{\mathscr{k}^{(a)}j} \times \pi^{*(a - 1)}_{\mathscr{k}^{(a - 1)}j} \times \dots \times \pi^{*(1)}_{\mathscr{k}^{(1)}j} \times \pi_{j} \Bigr).
\end{aligned}
\end{equation}
The contribution to the likelihood by a single subject $i$ is thus $\prod_{k^{(a)} = 1}^2 P(Y_i^{*(a)} = k^{(a)} | \boldsymbol{X_i}, \boldsymbol{Z^{(a)}_i}, \dots, \boldsymbol{Z^{(1)}_i})^{y^{*(a)}_{i k^{(a)}}}$ where $\pi^{*(a)}_{\mathscr{k}^{(a)}j} = P(Y^{*(a)} = k^{(a)} | \mathcal{Y}^{*(a - 1)}, Y = j, \boldsymbol{Z^{(a)}}, \dots, \boldsymbol{Z^{(1)}}; \boldsymbol{\gamma})$, $\pi_{j} = P(Y = j | \boldsymbol{X} ; \boldsymbol{\beta})$ and $y^{*(a)}_{i k^{(a)}} = \mathbb{I} \{ Y_i^{*(a)} = k^{(a)}\}$. We can estimate $(\boldsymbol{\beta}, \boldsymbol{\gamma})$ using the following observed data log-likelihood for subjects $i = 1, \dots, N$, 
\begin{equation}
\begin{aligned}
\label{eq:multistage-obs-log-like}
\ell_{obs}( \boldsymbol{\beta},  \boldsymbol{\gamma};  \boldsymbol{X},  \boldsymbol{Z^{(a)}}, \dots,  \boldsymbol{Z^{(1)}}) &\;
 = \sum_{i = 1}^N  \log \Bigl\{ \prod_{b = 1}^a \bigl( \prod_{k^{(a)} = 1}^2 P(Y_i^{*(b)} = k^{(b)} |  \boldsymbol{X_i}, \boldsymbol{Z^{(b)}_i}, \dots, \boldsymbol{Z^{(1)}_i})^{y^{*(b)}_{i k^{(b)}}} \bigr)  \Bigr\} \\
 & = \sum_{i = 1}^N \Bigl[ \sum_{b = 1}^a y^{*(b)}_{i k^{(b)}} \sum_{k^{(b)} = 1}^2 \log \{ P(Y_i^{*(b)} = k^{(b)} | \boldsymbol{X_i}, \boldsymbol{Z^{(b)}_i}, \dots, \boldsymbol{Z^{(1)}_i}) \} \Bigr] \\
 & = \sum_{i = 1}^N \Bigl[ \sum_{b = 1}^a y^{*(b)}_{i k^{(b)}} \sum_{k^{(b)} = 1}^2  \\
 &\qquad  \log \{ \sum_{j = 1}^2\sum_{k^{(b-1)}, \dots, k^{(1)} = 1}^2 \bigl( \pi^{*(b)}_{\mathscr{k}^{(b)}j} \times \pi^{*(b - 1)}_{\mathscr{k}^{(b - 1)}j} \times \dots \times \pi^{*(1)}_{\mathscr{k}^{(1)}j} \times \pi_{ij} \bigr) \} \Bigr]
\end{aligned}
\end{equation}
where $\sum_{k^{(a)}, \dots, k^{(1)} = 1}^2$ is equivalent to $\sum_{k^{(a)} = 1}^2, \sum_{k^{(a - 1)} =1}^2, \dots, \sum_{k^{(1)} = 1}^2$. As in the basic model, the observed data log-likelihood is difficult to use directly for optimization. Instead, we can view the true outcome $Y$ as a latent variable, and construct the complete data log-likelihood for a multistage model as follows,
\begin{equation}
\begin{aligned}
\label{eq:multistage-complete-log-like}
\ell_{complete}(\boldsymbol{\beta}, \boldsymbol{\gamma}; \boldsymbol{X}, \boldsymbol{Z^{(a)}},\dots, \boldsymbol{Z^{(1)}}) &\; = 
 \sum_{i = 1}^N \Bigl[ \sum_{j = 1}^2 y_{ij} \text{log} \{ P(Y_i = j | \boldsymbol{X_i}) \} &\; \\
  &\qquad + \sum_{b = 1}^a \Bigl( \sum_{j = 1}^2 \sum_{k^{(a)}, \dots, k^{(1)} = 1}^2 y_{ij} \times \\
 &\qquad \bigl(\prod_{h = 1}^b y^{*(h)}_{ik^{(h)}} \bigr) \text{log} \{ P(Y^{*(b)}_i = k^{(b)} |\mathcal{Y}^{*(b - 1)}, Y_i = j, \boldsymbol{Z^{(b)}}) \} \Bigr) \Bigr] \\
  & = \sum_{i = 1}^N \Bigl[ \sum_{j = 1}^2 y_{ij} \text{log} \{ \pi_{ij} \} \\
  &\qquad+\sum_{b = 1}^a \Bigl( \sum_{j = 1}^2 \sum_{k^{(a)}, \dots, k^{(1)} = 1}^2 y_{ij} \bigl(\prod_{h = 1}^b y^{*(h)}_{ik^{(h)}} \bigr) \text{log} \{ \pi^{*(b)}_{i \mathscr{k}^{(b)} j} \} \Bigr) \Bigr].
\end{aligned}
\end{equation}
Since we do not observe the true outcome value $Y$, the complete data log-likelihood cannot be used for maximization. Note that (\ref{eq:multistage-complete-log-like}) can be viewed as a mixture model with latent mixture components, $y_{ij}$, and covariate-dependent mixing proportions $\pi_{ij}$.

\section{Details on Estimation Methods}\label{estimation-details}
In this section, we provide a detailed description of our estimation methods for the proposed multistage misclassification model. First, we present an Expectation-Maximization algorithm to jointly estimate $\boldsymbol{\beta}$ and $\boldsymbol{\gamma}$ \citep{dempster1977maximum}. Next, we propose Bayesian methods for this setting. Both estimation strategies are available in the R Package \textit{COMBO} for a two-stage misclassification model \citep{combo}.

\subsection{Maximization Using an EM Algorithm}\label{estimation-details-em}
For the EM algorithm, we begin with the complete data log-likelihood defined in (\ref{eq:multistage-complete-log-like}). Because (\ref{eq:multistage-complete-log-like}) is linear in $y_{ij}$, we can replace $y_{ij}$ with the following quantity in the E-step of the algorithm,
\begin{equation}
\begin{aligned}
\label{eq:multistage-e-step}
w^{(a)}_{ij} = P(Y_i = j | Y_i^{*(a)}, \dots, Y_i^{*(1)}, \boldsymbol{X}, \boldsymbol{Z^{(a)}}, \dots, \boldsymbol{Z^{(1)}}) = \sum_{k^{(a)}, \dots, k^{(1)}= 1}^2 \frac{\pi_{ij} \bigl(  \prod_{b = 1}^a y^{*(b)}_{ik^{(b)}} \pi^{*(b)}_{i \mathscr{k}^{(b)} j} \bigr) }{\sum_{\ell = 1}^2  \pi_{i \ell}\bigl(  \prod_{b = 1}^a \pi^{*(b)}_{i \mathscr{k}^{(b)} \ell } \bigr) }. 
\end{aligned}
\end{equation}
In the M-step, we maximize the following expected log-likelihood with respect to $\boldsymbol{\beta}$ and $\boldsymbol{\gamma}$,
\begin{equation}
\begin{aligned}
\label{eq:multistage-m-step}
Q = \sum_{i = 1}^N \Bigl[ \sum_{j = 1}^2 w^{(a)}_{ij} \text{log} \{ \pi_{ij} \} + 
  \sum_{b = 1}^a \Bigl( \sum_{j = 1}^2 \sum_{k^{(a)}, \dots, k^{(1)} = 1}^2 w^{(a)}_{ij} \bigl(\prod_{h = 1}^b y^{*(h)}_{ik^{(h)}} \bigr) \text{log} \{ \pi^{*(b)}_{i \mathscr{k}^{(b)} j} \} \Bigr) \Bigr].
\end{aligned}
\end{equation}
After estimates for $\boldsymbol{\beta}$ and $\boldsymbol{\gamma}$ are obtained from this EM algorithm, Algorithm \ref{alg:multistage-label-switch} must be used to correct potential label switching and return final parameter estimates. The covariance matrix for $\beta$ and $\gamma$ is obtained by inverting the expected information matrix. The covariance matrix for $\boldsymbol{\beta}$ and $\boldsymbol{\gamma}$ is obtained by inverting the expected information matrix, with relabeling requirements as described in \cite{hochstedler2023statistical}.

\subsection{Bayesian Modeling}\label{estimation-details-mcmc}
Our proposed multistage misclassification model is defined for each of the $a$ stages in the model: $Y^{* (a)}_i | \pi^{* (a)}_i \sim Bernoulli ( \pi^{* (a)}_i)$. Here, $\pi^{* (a)}_i = \sum_{j = 1}^2 \sum_{k^{(a - 1)}, \dots, k^{(1)} = 1}^{2} \pi^{* (a)}_{i \mathscr{k}^{(a)} j} \times \pi^{* (a - 1)}_{i \mathscr{k}^{(a - 1)} j} \times \dots \times \pi^{* (1)}_{i \mathscr{k}^{(1)} j} \times \pi_{ij}$. We estimate this model using a Markov Chain Monte Carlo (MCMC) procedure. Prior distributions for the parameters should be determined using input from subject-matter experts, based on the context of the problem that the model is applied to. We recommend proper, relatively flat priors. In the R Package, \textit{COMBO}, users can specify prior parameters for either Uniform, Normal, Double Exponential, or t prior distributions. Before summarizing the results, Algorithm \ref{alg:multistage-label-switch} must be applied on \textit{each individual MCMC chain} to correct for label switching, if it is present. Standard methods are used to compute variance metrics. 

\section{Label Switching}\label{label-switching}
Since the complete data log-likelihood in (\ref{eq:multistage-complete-log-like}) for the multistage model is a mixture likelihood, it is also invariant under relabeling of the mixture components, $y_{ij}$. Regardless of the number of stages in the model, $a$, there are two mixture components and therefore $J! = 2! = 2$ plausible parameter sets \citep{betancourt2017identifying, stephens2000dealing}. In addition, the pattern that governs these parameter sets is identical to that described in \cite{hochstedler2023statistical}, the $\boldsymbol{\beta}$ parameters change signs and the $\boldsymbol{\gamma}$ parameters change $j$ subscripts. Once again, to correct for label switching we compute Youden's $J$ Statistic for a given observation stage $a$ \citep{BERRAR2019546}, 
\begin{equation}
\begin{aligned}
\label{eq:paper2-J-stats}
&\hat{J}^{(a)} = \hat{\pi}^{*(a)}_{11} + \hat{\pi}^{*(a)}_{22} - 1,\\
&\hat{J}^{(a), switch} = \hat{\pi}^{*(a), switch}_{11} + \hat{\pi}^{*(a), switch}_{22} - 1.
\end{aligned}
\end{equation}
In (\ref{eq:paper2-J-stats}), we define $\hat{\pi}^{*(a)}_{jj}$ as follows:
\begin{equation}
\begin{aligned}
\label{eq:paper2-avg_pistar}
\hat{\pi}^{*(a)}_{11} = \frac{1}{N}\sum_{i = 1}^N \hat{\pi}^{*(a)}_{i11} = \frac{1}{N}\sum_{i = 1}^N \frac{\text{exp}\{\hat{\gamma}^{(a)}_{110} + \boldsymbol{\hat{\gamma}^{(a)}_{11Z} Z^{(a)}_i}\}}{1 + \text{exp}\{\hat{\gamma}^{(a)}_{110} + \boldsymbol{\hat{\gamma}^{(a)}_{11Z} Z^{(a)}_i}\}},\\
\hat{\pi}^{*(a)}_{22} = \frac{1}{N}\sum_{i = 1}^N \hat{\pi}^{*(a)}_{i22} = \frac{1}{N}\sum_{i = 1}^N \frac{1}{1 + \text{exp}\{\hat{\gamma}^{(a)}_{120} + \boldsymbol{\hat{\gamma}^{(a)}_{12Z} Z^{(a)}_i}\}}.
\end{aligned}
\end{equation}
That is, the average sensitivity and specificity estimates respectively, $\hat{\pi}^{*(a)}_{11}$ and $\hat{\pi}^{*(a)}_{22}$, are computed for a given outcome stage using the parametric form described in (\ref{eq:appendix-multistage_conceptual_framework_eq}) with parameter estimates from any of the proposed estimation strategies in Section \ref{estimation-details}. We obtain $\hat{\pi}^{*(a), switch}_{11}$ and $\hat{\pi}^{*(a), switch}_{22}$ through the following relationship:
\begin{equation}
\begin{aligned}
\label{eq:paper2-misclassification_label_switch}
\hat{\pi}^{*(a), switch}_{11} = 1 - \hat{\pi}^{*(a)}_{22},\\
\hat{\pi}^{*(a), switch}_{22} = 1 - \hat{\pi}^{*(a)}_{11}.
\end{aligned}
\end{equation}
A procedure to correct label switching in a multistage misclassification model is provided in Algorithm \ref{alg:multistage-label-switch} \citep{hochstedler2023statistical}.
\begin{algorithm}
\caption{Correcting label switching in multistage binary outcome misclassification models}\label{alg:multistage-label-switch}
\begin{algorithmic}
\State Compute $\hat{J}^{(a)}$ and $\hat{J}^{(a), switch}$ for stage $a$ using $\hat{\boldsymbol{\beta}}$ and $\hat{\boldsymbol{\gamma}}$ and perform the following for all outcome stages $b \in \{1, \dots, a \}$.
\If{$\hat{J}^{(a)} \geq \hat{J}^{(a), switch}$}
    \State $\hat{\beta}_{corrected} \gets \hat{\beta}$
    \State $\hat{\gamma}_{corrected} \gets \hat{\gamma}$
\Else
    \State $\hat{\beta}_{corrected} \gets -\hat{\beta}$
    \State $\hat{\gamma}^{(b)}_{corrected, \mathscr{k}^{(b)} 1} \gets \hat{\gamma}^{(b)}_{\mathscr{k}^{(b)} 2}$
    \State $\hat{\gamma}^{(b)}_{corrected, \mathscr{k}^{(b)} 2} \gets \hat{\gamma}^{(b)}_{\mathscr{k}^{(b)} 1}$
\EndIf 
\end{algorithmic}
\end{algorithm}

Note that Algorithm \ref{alg:multistage-label-switch} uses average conditional response probabilities for a single observation stage $a$. Depending on the problem context, analysts may instead choose to implement Algorithm \ref{alg:multistage-label-switch} using average conditional response probabilities from multiple model stages. This choice would create a more strict criterion for label switching, but would ensure sensible sensitivity and specificity estimates at more model stages. 

\section{Simulation Studies} \label{appendix-a}

\subsection{Simulation Settings} 
We present simulations for evaluating the proposed binary outcome misclassification model in terms of bias and root mean squared error (rMSE) for two-stage cases. For a given simulation scenario, we present parameter estimates for a binary outcome misclassification model obtained from the EM-algorithm and from MCMC, under a \textit{Uniform}$(-10,10)$ prior distribution setting. We compare our estimates to a naive \textit{analysis model} that assumes $Y^{*(1)}$ and $Y^{*(2)}$ are measured without error. 

In all settings, we generate $500$ datasets with $P(Y = 1) \approx 65\%$. In the simulation settings for the two-stage model, we consider four simulation scenarios. First, we examine the case of a relatively small sample size and high misclassification rate, as this case would likely be highly problematic in a two-stage example. In this setting, datasets had $1000$ members and the imposed outcome misclassification rates for $Y^{*(1)}$ were between $10\%$ and $22\%$. The probability of correct measurement across all stages was set between $79\%$ and $92\%$. In Setting 2, we show that even with two sequential observed outcomes and a large sample size, even small misclassification rates can still impact parameter estimation. In this setting, we generated datasets with $10000$ members and imposed misclassification rates in $Y^{*(1)}$ between $4\%$ and $9\%$. In the third simulation setting, we evaluated the case where $Y^{*(1)}$ is measured without error, but $Y^{*(2)}$ is subject to misclassification. In Setting 4, we instead consider the case where $Y^{*(1)}$ is subject to misclassification, but $Y^{*(2)}$ has perfect specificity. These scenarios demonstrate that our multistage methods are still appropriate for cases where there is perfect measurement in at least one of the observed outcomes. In both Setting 3 and Setting 4, we generated datasets with $1000$ members and imposed misclassification rates between $10\%$ and $20\%$ in the imperfectly measured outcome.

For a dataset with $1,000$ members, the two-stage analysis using our proposed EM algorithm took about $25$ seconds. The MCMC analysis took considerably longer, at approximately $3.5$ hours. 

These settings are outlined in Table \ref{sim-setting-table2}. All analyses were conducted in \texttt{R} \citep{stats2021R}.

\begin{table}[H]
\centering
\caption{Number of generated datasets (N. Realizations), Sample size ($N$), $P(Y = 1)$, first-stage sensitivity ($P(Y^* = 1 | Y = 1)$), first-stage specificity ($P(Y^* = 2 | Y = 2)$), second-stage sensitivity ($P(Y^{*(2)} = 1 | Y^{*(1)} = 1, Y = 1)$), second-stage specificity ($P(Y^{*(2)} = 2 | Y^{*(1)} = 2, Y = 2)$), $\beta$ prior distribution, and $\gamma$ prior distribution settings for each of the the simulation Settings 4, 5, and 6.} \label{sim-setting-table2}
\begin{threeparttable}
\begin{tabular}{clrrrrr}
\hline
        Scenario & & & & Setting \\
\hline
    \\
(1)           && N. Realizations & & 500\\
              && $N$ & & 1000\\
              && $P(Y = 1)$ & & 0.65 \\
              && $P(Y^{*(1)} = 1 | Y = 1)$ && 0.83 - 0.91 \\
              && $P(Y^{*(1)} = 2 | Y = 2)$ && 0.79 - 0.85\\
              && $P(Y^{*(2)} = 1 | Y^{*(1)} = 1, Y = 1)$ && 0.85 - 0.92 \\
              && $P(Y^{*(2)} = 2 | Y^{*(1)} = 2, Y = 2)$ && 0.79 - 0.87\\
              && $\beta$ prior distribution && \textit{Uniform}$(-10, 10)$ \\
              && $\gamma$ prior distribution && \textit{Uniform}$(-10, 10)$ \\
              \\
(2)           && N. Realizations & & 500\\
              && $N$ & & 10000\\
              && $P(Y = 1)$ & & 0.65 \\
              && $P(Y^{*(1)} = 1 | Y = 1)$ && 0.91 - 0.93 \\
              && $P(Y^{*(1)} = 2 | Y = 2)$ && 0.91 - 0.93 \\
              && $P(Y^{*(2)} = 1 | Y^{*(1)} = 1, Y = 1)$ && 0.94 - 0.96 \\
              && $P(Y^{*(2)} = 2 | Y^{*(1)} = 2, Y = 2)$ && 0.91 - 0.93\\
              && $\beta$ prior distribution && \textit{Uniform}$(-10, 10)$ \\
              && $\gamma$ prior distribution && \textit{Uniform}$(-10, 10)$ \\
              \\
(3)           && N. Realizations & & 500\\
              && $N$ & & 1000\\
              && $P(Y = 1)$ & & 0.65 \\
              && $P(Y^{*(1)} = 1 | Y = 1)$ && 1 \\
              && $P(Y^{*(1)} = 2 | Y = 2)$ && 1\\
              && $P(Y^{*(2)} = 1 | Y^{*(1)} = 1, Y = 1)$ && 0.88 - 0.92\\
              && $P(Y^{*(2)} = 2 | Y^{*(1)} = 2, Y = 2)$ && 0.82 - 0.90\\
              && $\beta$ prior distribution && \textit{Uniform}$(-10, 10)$ \\
              && $\gamma$ prior distribution && \textit{Uniform}$(-10, 10)$ \\
              \\
(4)           && N. Realizations & & 500\\
              && $N$ & & 1000\\
              && $P(Y = 1)$ & & 0.65 \\
              && $P(Y^{*(1)} = 1 | Y = 1)$ && 0.84 - 0.90 \\
              && $P(Y^{*(1)} = 2 | Y = 2)$ && 0.80 - 0.89 \\
              && $P(Y^{*(2)} = 1 | Y^{*(1)} = 1, Y = 1)$ && 0.88 - 0.92 \\
              && $P(Y^{*(2)} = 2 | Y^{*(1)} = 2, Y = 2)$ && 1\\
              && $\beta$ prior distribution && \textit{Uniform}$(-10, 10)$ \\
              && $\gamma$ prior distribution && \textit{Uniform}$(-10, 10)$ \\
\hline  
\end{tabular}
\end{threeparttable}
\end{table}

\subsection{Data Generation}
For each of the two-stage simulated datasets, we begin by generating the predictor $X$ from a standard Normal distribution and the predictors $Z^{(1)}$ and $Z^{(2)}$ from a Gamma distribution. The shape parameters for the Gamma distributions were 1 for both $Z^{(1)}$ and $Z^{(2)}$ in Setting 1, Setting 3, and Setting 4. In Setting 2, the shape parameters for the Gamma distributions were 2 for both $Z^{(1)}$ and $Z^{(2)}$. For Settings 1-4, we used the following relationship to generate the true outcome status: $P(Y = 1 | X) = 1 + (-2)X$. For Setting 1, Setting 2, and Setting 4, we obtained $Y^{*(1)}$ using the following relationships: $P(Y^{*(1)} = 1 | Y = 1, Z^{(1)}) = 1 + (1)Z^{(1)}$ and $P(Y^{*(1)} = 1 | Y = 2, Z^{(1)}) = -0.50 + (-1.5)Z^{(1)}$. In Setting 3, $P(Y^{*(1)} = 1 | Y = 1, Z^{(1)}) = 5 + (5)Z^{(1)}$ and $P(Y^{*(1)} = 1 | Y = 2, Z^{(1)}) = -5 + (-5)Z^{(1)}$. The choice of parameter values resulted in near perfect sensitivity and specificity for $Y^{*(1)}$ in the generated datasets. For Settings 1-3, we obtained $Y^{*(2)}$ using the following relationships: $P(Y^{*(2)} = 1 | Y^{*(1)} = 1, Y = 1, Z^{(2)}) = 1.5 + (1)Z^{(2)}$,
$P(Y^{*(2)} = 1 | Y^{*(1)} = 2, Y = 1, Z^{(2)}) = 0.50 + (0.50)Z^{(2)}$, $P(Y^{*(2)} = 1 | Y^{*(1)} = 1, Y = 2, Z^{(2)}) = -0.50 + (0)Z^{(2)}$, and $P(Y^{*(2)} = 1 | Y^{*(1)} = 2, Y = 2, Z^{(2)}) = -1 + (-1)Z^{(2)}$. In Setting 4, the same relationships for $P(Y^{*(2)} = 1 | Y^{*(1)} = 1, Y = 1, Z^{(2)})$ and
$P(Y^{*(2)} = 1 | Y^{*(1)} = 2, Y = 1, Z^{(2)})$ were retained, but $P(Y^{*(2)} = 1 | Y^{*(1)} = 2, Y = 2, Z^{(2)}) = P(Y^{*(2)} = 1 | Y^{*(1)} = 1, Y = 2, Z^{(2)}) = -5 + (-5)Z^{(2)}$, ensuring perfect specificity in $Y^{*(2)}$.

\subsection{Simulation Study Results} \label{simulations}
We present simulations for evaluating the proposed binary outcome misclassification model in terms of bias and root mean squared error (rMSE) for two-stage cases. The two-stage mislcassification model is compared to a naive two-stage model that assumes no measurement error in \textit{both} of the observed outcomes. 

We investigate four simulation settings with varying sample sizes and misclassification rates: (1) small sample size and large misclassification rates and (2) large sample size and small misclassification rates, (3) small sample size and perfect $Y^{*(1)}$ and, (4) small sample size and perfect specificity in $Y^{*(2)}$. We present results from the EM algorithm for settings 1-4 and results from MCMC for settings 1, 3, and 4. MCMC was not performed for setting 2 due to computational time constraints (see Appendix \ref{appendix-a}). Details on these settings can be found in Appendix \ref{appendix-a}.  

Table \ref{parameter-results-table-setting4-5} and Table \ref{parameter-results-table-setting6-7} present mean parameter estimates and rMSE across 500 simulated datasets for simulation settings 1-4. For each simulation setting, Table \ref{probability-results-table-2} presents the true outcome probability, first-stage sensitivity, first-stage specificity, second-stage sensitivity, and second-stage specificity values measured from the generated data and estimated from the EM algorithm and MCMC results.

\begin{table}[h!]\footnotesize
\centering
\caption{Bias and root mean squared error (rMSE) for parameter estimates from 500 realizations of simulation Settings 1 and 2. ``EM'' and ``MCMC'' estimates were computed using the \textit{COMBO} R Package. The ``Naive Analysis'' results were obtained by a two-stage model that does not account for outcome misclassification. Estimates marked with a ``-'' are not obtained by the given estimation method.} \label{parameter-results-table-setting4-5}
\begin{threeparttable}
\begin{tabular}{clrrrrrrrrrrrr}
\hline
        &       & \multicolumn{2}{c}{EM}       &
        \multicolumn{2}{c}{MCMC}&  
        \multicolumn{2}{c}{Naive Analysis}\\
        \cline{3-8}
Scenario &       & \multicolumn{1}{c}{Bias} & \multicolumn{1}{c}{rMSE} & \multicolumn{1}{c}{Bias} & \multicolumn{1}{c}{rMSE} & \multicolumn{1}{c}{Bias} & \multicolumn{1}{c}{rMSE} \\
\hline
    \\
      & $\beta_0$ & 0.032 & 0.197 & -0.053 & 0.224 & 0.029 & 0.299 \\
(1)   & $\beta_X$ & -0.101 & 0.301 & 0.261 & 0.490 & -0.100 & 0.547 \\
      & $\gamma^{(1)}_{110}$ & -0.028 & 0.209 & 0.249 & 0.522 & - & - \\
      & $\gamma^{(1)}_{11Z^{(1)}}$ & 0.044 & 0.338 & 0.405 & 1.046 & - & - \\
      & $\gamma^{(1)}_{120}$ & 0.067 & 0.333 & -0.496 & 1.210 & - & - \\
      & $\gamma^{(1)}_{12Z^{(1)}}$ & -0.156 & 0.656 & -1.425 & 2.168 & - & - \\
      & $\gamma^{(2)}_{1110}$ & -0.014 & 0.249 & 1.697 & 2.385 & -0.195 & 0.364 \\
      & $\gamma^{(2)}_{111Z^{(2)}}$ & 0.047 & 0.374 & 2.674 & 2.989 & 0.114 & 1.233 \\
      & $\gamma^{(2)}_{1210}$ & -0.022 & 0.652 & -2.938 & 4.341 & - & - \\
      & $\gamma^{(2)}_{121Z^{(2)}}$ & 0.128 & 0.710 & 2.674 & 3.664 & - & - \\
      & $\gamma^{(2)}_{1120}$ & 0.017 & 1.236 & 2.387 & 3.925 & - & -\\
      & $\gamma^{(2)}_{112Z^{(2)}}$ & -0.398 & 3.091 & -0.124 & 3.091 & - & -\\
      & $\gamma^{(2)}_{1220}$ & 0.026 & 0.343 & -1.918 & 2.647 & -0.073 & 3.909 \\
      & $\gamma^{(2)}_{122Z^{(2)}}$ & -0.125 & 0.581 & -3.025 & 3.247 & -0.390 & 3.251 \\
      \\
      & $\beta_0$ & 0.003 & 0.036 &  &  & 0.006 & 0.058 \\
(2)   & $\beta_X$ & -0.005 & 0.062 &  &  & 0.026 & 0.101 \\
      & $\gamma^{(1)}_{110}$ & -0.011 & 0.108 & & & - & - \\
      & $\gamma^{(1)}_{11Z^{(1)}}$ & 0.013 & 0.096 & & & - & - \\
      & $\gamma^{(1)}_{120}$ & 0.036 & 0.188 && & - & - \\
      & $\gamma^{(1)}_{12Z^{(1)}}$ & -0.035 & 0.213 & & & - & - \\
      & $\gamma^{(2)}_{1110}$ & -0.012 & 0.111 & & & -0.053 & 0.157 \\
      & $\gamma^{(2)}_{111Z^{(2)}}$ & 0.013 & 0.108 & & & -0.128 & 0.208 \\
      & $\gamma^{(2)}_{1210}$ & 0.166 & 0.413 && & - & - \\
      & $\gamma^{(2)}_{121Z^{(2)}}$ & -0.061 & 0.211 & & & - & - \\
      & $\gamma^{(2)}_{1120}$ & -0.110 & 0.492 & & & - & -\\
      & $\gamma^{(2)}_{112Z^{(2)}}$ & 0.024 & 0.234 && & - & -\\
      & $\gamma^{(2)}_{1220}$ & -0.027 & 0.151 && & -0.092 & 0.239 \\
      & $\gamma^{(2)}_{122Z^{(2)}}$ & 0.006 & 0.133 & & & 0.234 & 0.363 \\
      \\
\hline  
\end{tabular}
\end{threeparttable}
\end{table}

\begin{table}[h!]\footnotesize
\centering
\caption{Bias and root mean squared error (rMSE) for parameter estimates from 500 realizations of simulation Settings 3 and 4. ``EM'' and ``MCMC'' estimates were computed using the \textit{COMBO} R Package. The ``Naive Analysis'' results were obtained by a two-stage model that does not account for outcome misclassification. Estimates marked with a ``-'' are not obtained by the given estimation method.} \label{parameter-results-table-setting6-7}
\begin{threeparttable}
\begin{tabular}{clrrrrrrrrrrrr}
\hline
        &       & \multicolumn{2}{c}{EM}       &
        \multicolumn{2}{c}{MCMC}&  
        \multicolumn{2}{c}{Naive Analysis}\\
        \cline{3-8}
Scenario &       & \multicolumn{1}{c}{Bias} & \multicolumn{1}{c}{rMSE} & \multicolumn{1}{c}{Bias} & \multicolumn{1}{c}{rMSE} & \multicolumn{1}{c}{Bias} & \multicolumn{1}{c}{rMSE} \\
\hline
      \\
      & $\beta_0$ & -0.018 & 0.118 & -0.001 & 0.113 & 0.016 & 0.247\\
(3)   & $\beta_X$ & -0.097 & 0.195 & 0.013 & 0.166 & -0.131 & 0.457 \\
      & $\gamma_{110}$ & 19.373 & 65.503 & 1.26 & 1.640 & - & - \\
      & $\gamma_{11Z^{(1)}}$ & -4.068 & 14.446 & -0.450 & 1.094 & - & - \\
      & $\gamma_{120}$ & -43.863 & 186.594 & -0.815 & 1.397 & - & - \\
      & $\gamma_{12Z^{(1)}}$ & -4.388 & 285.955 & 0.678 & 1.295 & - & - \\
      & $\gamma^{(2)}_{1110}$ & -0.016 & 0.194 & -0.033 & 0.327 & -0.034 & 0.325 \\
      & $\gamma^{(2)}_{111Z^{(2)}}$ & 0.064 & 0.282 & 0.527 & 0.969 & 0.239 & 0.898 \\
      & $\gamma^{(2)}_{1210}$ & 1.247 & 2.059 & -0.797 & 1.084 & - & - \\
      & $\gamma^{(2)}_{121Z^{(2)}}$ & 1.224 & 2.035 & -0.742 & 1.049 & - & - \\
      & $\gamma^{(2)}_{1120}$ & 2.191 & 2.820 & 0.872 & 1.236 & - & - \\
      & $\gamma^{(2)}_{112Z^{(2)}}$ & 1.727 & 2.242 & 0.345 & 0.883 & - & -\\
      & $\gamma^{(2)}_{1220}$ & 0.000 & 0.228 & -0.006 & 0.631 & -0.024 & 2.788 \\
      & $\gamma^{(2)}_{122Z^{(2)}}$ & -0.082 & 0.335 & -1.019 & 1.589 & -0.968 & 6.350 \\
              \\
      & $\beta_0$ & 0.016 & 0.165 & -0.049 & 0.230 & 0.006 & 0.232\\
(4)   & $\beta_X$ & -0.160 & 0.266 & 0.037 & 0.272 & -0.149 & 0.349 \\
      & $\gamma_{110}$ & -0.057 & 0.197 & 0.101 & 0.293 & - & -\\
      & $\gamma_{11Z^{(1)}}$ & 0.061 & 0.264 & 0.211 & 0.749 & - & -\\
      & $\gamma_{120}$ & 0.133 & 0.282 & 0.017 & 0.458 & - & -\\
      & $\gamma_{12Z^{(1)}}$ & -0.042 & 0.440 & -0.655 & 1.292 & - & -\\
      & $\gamma^{(2)}_{1110}$ & 0.011 & 0.252 & 1.119 & 1.823 & -0.195 & 0.349\\
      & $\gamma^{(2)}_{111Z^{(2)}}$ & 0.034 & 0.394 & 1.888 & 2.381 & 0.033 & 1.549\\
      & $\gamma^{(2)}_{1210}$ & -0.114 & 0.509 & -2.304 & 4.178 & - & -\\
      & $\gamma^{(2)}_{121Z^{(2)}}$ & 0.004 & 0.461 & 2.610 & 3.502 & - & -\\
      & $\gamma^{(2)}_{1120}$ & -0.067 & 9.275 & -0.134 & 1.197 & - & -\\
      & $\gamma^{(2)}_{112Z^{(2)}}$ & 1.251 & 8.798 & 1.370 & 2.132 & - & -\\
      & $\gamma^{(2)}_{1220}$ & -12.413 & 29.590 & -1.238 & 1.365 & -4.833 & 18.441\\
      & $\gamma^{(2)}_{122Z^{(2)}}$ & -9.996 & 24.390 & 0.413 & 0.731 & 3.482 & 10.316\\
              \\
\hline  
\end{tabular}
\end{threeparttable}
\end{table}

\begin{table}[h!]
\centering
\caption{Estimated event probabilities from 500 realizations of simulation Settings 1-4. ``Data'' terms refer to empirical values computed from generated datasets. ``EM'' and ``MCMC'' estimates were computed using the \textit{COMBO} R Package.} \label{probability-results-table-2}
\begin{threeparttable}
\begin{tabular}{clrrrrr}
\hline
        Scenario & & & Data & EM & MCMC \\
\hline
    \\
(1)           && $P(Y = 1)$ & 0.648 & 0.647 & 0.650 \\
              && $P(Y = 2)$ & 0.352 & 0.353 & 0.350 \\
              && $P(Y^{*(1)} = 1 | Y = 1)$ & 0.852 & 0.848 & 0.888 \\
              && $P(Y^{*(1)} = 2 | Y = 2)$ & 0.822 & 0.816 & 0.897 \\
              && $P(Y^{*(2)} = 1 | Y^{*(1)} = 1, Y = 1)$ & 0.903 & 0.901 & 0.978 \\
              && $P(Y^{*(2)} = 2 | Y^{*(1)} = 2, Y = 2)$ & 0.853 & 0.851 & 0.973 
              \\
              \\
              \\
(2)           && $P(Y = 1)$ & 0.648 & 0.648 &  \\
              && $P(Y = 2)$ & 0.352 & 0.352 &  \\
              && $P(Y^{*(1)} = 1 | Y = 1)$ & 0.921 & 0.920 &  \\
              && $P(Y^{*(1)} = 2 | Y = 2)$ & 0.921 & 0.919 &  \\
              && $P(Y^{*(2)} = 1 | Y^{*(1)} = 1, Y = 1)$ & 0.949 & 0.949 & \\
              && $P(Y^{*(2)} = 2 | Y^{*(1)} = 2, Y = 2)$ & 0.920 & 0.921 & \\
              \\
              \\
(3)           && $P(Y = 1)$ & 0.647 & 0.640 & 0.648 \\
              && $P(Y = 2)$ & 0.353 & 0.360 & 0.352 \\
              && $P(Y^{*(1)} = 1 | Y = 1)$ & 1 & 0.997 & 0.999 \\
              && $P(Y^{*(1)} = 2 | Y = 2)$ & 1 & 0.977 & 0.999 \\
              && $P(Y^{*(2)} = 1 | Y^{*(1)} = 1, Y = 1)$ & 0.902 & 0.903 & 0.917 \\
              && $P(Y^{*(2)} = 2 | Y^{*(1)} = 2, Y = 2)$ & 0.851 & 0.855 & 0.890 \\
              \\
              \\
(4)           && $P(Y = 1)$ & 0.648 & 0.642 & 0.641 \\
              && $P(Y = 2)$ & 0.352 & 0.358 & 0.359 \\
              && $P(Y^{*(1)} = 1 | Y = 1)$ & 0.851 & 0.847 & 0.871 \\
              && $P(Y^{*(1)} = 2 | Y = 2)$ & 0.820 & 0.803 & 0.848 \\
              && $P(Y^{*(2)} = 1 | Y^{*(1)} = 1, Y = 1)$ & 0.903 & 0.903 & 0.967 \\
              && $P(Y^{*(2)} = 2 | Y^{*(1)} = 2, Y = 2)$ & 1 & 0.994 & 0.999 \\
              \\
\hline  
\end{tabular}
\end{threeparttable}
\end{table}

\textbf{Setting 1:} Across all simulated datasets, the average $P(Y = 1)$ was $64.8\%$, the average $P(Y^{*(1)} = 1)$ was $61.5\%$, and the average $P(Y^{*(2)} = 1)$ was $63.3\%$. The average first-stage correct classification rate was $85.2\%$ for $Y^{*(1)} = 1$ and $82.2\%$ for $Y^{*(1)} = 2$. The average second-stage correct classification rate was $90.3\%$ for $Y^{*(2)} = 1$ and $85.3\%$ for $Y^{*(2)} = 2$ (Table \ref{probability-results-table-2}). The naive analysis results in small bias for $\beta_X$, but the rMSE is large compared to the EM Algorithm and MCMC methods that account for potential misclassification (Table \ref{parameter-results-table-setting4-5}). Our proposed EM algorithm performs well for the $\boldsymbol{\beta}$ parameters and first-stage $\boldsymbol{\gamma}$ parameters. Some of the second-stage $\boldsymbol{\gamma}$ parameter estimates have wider variation. While our proposed MCMC method performs well for the $\boldsymbol{\beta}$ parameter estimates, the bias and rMSE for the $\boldsymbol{\gamma}$ estimates are considerably higher than that from the EM algorithm. Both the EM algorithm and MCMC methods recover the true outcome probabilities $P(Y = 1)$ and $P(Y = 2)$ in Table \ref{probability-results-table-2}, but MCMC tends to consistently overestimate correct classification rates for both stages of model. 

\textbf{Setting 2:} In Setting 2, the average $P(Y = 1)$ was $64.8\%$ (Table \ref{probability-results-table-2}). The observed outcome response probabilities were $62.4\%$ for $Y^{*(1)}$ and $64.4\%$ for $Y^{*(2)}$. The average first-stage sensitivity and specificity were both $92.1\%$. For second-stage outcomes, the average correct classification rate was  $94.9\%$ for $Y^{*(2)} = 1$ and $92.0\%$ for $Y^{*(2)} = 2$.While the bias for naive $\boldsymbol{\beta}$ estimates is not large, the naive method rMSE is higher than that of the two-stage misclassification model (Table \ref{parameter-results-table-setting4-5}). The two-stage misclassification model achieves low bias across all parameter estimates using the EM Algorithm. In addition the EM Algorithm results in near-perfect recovery of response probabilities and conditional response probabilities for $Y$, $Y^{*(1)}$, and $Y^{*(2)}$ (Table \ref{probability-results-table-2}). It should be noted that there were numerical issues in the estimation of many realizations for Setting 2. Scaling this method to large sample sizes is a topic under investigation. 

\textbf{Setting 3:} In Setting 3, the average response probabilities for each marginal outcome, $Y$, $Y^{*(1)}$, and $Y^{*(2)}$ were $64.7\%$, $64.7\%$, and $63.6\%$, respectively. Per the simulation design, $Y^{*(1)}$ was measured without error. The second-stage correct classification rate was $90.2\%$ for $Y^{*(2)} = 1$ and $85.1\%$ for $Y^{*(2)} = 2$ (Table \ref{probability-results-table-2}). In this setting, the naive model is an appropriate choice for the data.  As such, we find low bias in the naive $\boldsymbol{\beta}$ estimates in Table \ref{parameter-results-table-setting6-7}. Across both the EM Algorithm and MCMC estimates, we find substantial bias and large rMSE estimates for the first-stage $\boldsymbol{\gamma}$ parameters. This is not concerning because the first-stage $\boldsymbol{\gamma}$ parameters govern the misclassification mechanism for $Y^{*(1)}$, and there is no first-stage misclassification in the design. Similarly, the second-stage $\boldsymbol{\gamma}$ parameters associated with mismatched $Y$ and $Y^{*(1)}$ terms (i.e. $Y = 1$ and $Y^{*(1)} = 2$) are estimated with considerable bias using both the EM algorithm and MCMC. Since mismatched $Y$ and $Y^{*(1)}$ terms are not possible in this study design, it is unsurprising that the corresponding regression parameters are estimated poorly by the proposed methods. The remaining second-stage $\boldsymbol{\gamma}$ terms, however, are estimated with low bias and with rMSE well under the naive model using the EM algorithm. MCMC also produces reasonable estimates, but the bias and rMSE are both higher than that of the EM algorithm. In Table \ref{probability-results-table-2}, we see that both the EM and MCMC methods provide accurate estimates of $Y$ response probabilities. Importantly, both methods also correctly capture the perfect measurement of $Y^{*(1)}$, with correct classification probabilities estimated between $97.7\%$ and $99.9\%$. Second-stage correct classification rates are also estimated accurately using the EM Algorithm. These probabilities are slightly overestimated using MCMC. 

\textbf{Setting 4:} In Setting 4, the average response probability for $Y = 1$ was $64.8\%$ (Table \ref{probability-results-table-2}). The average response probabilities for the first-stage and second-stage observed outcomes were $61.5\%$ and $56.7\%$, respectively. The average first-stage correct classification rate was $85.1\%$ for $Y^{*(1)} = 1$ and $82.0\%$ for $Y^{*(1)} = 2$. The average second-stage correct classification rate was $90.3\%$ for $Y^{*(2)} = 1$. Per the simulation design, $P(Y^{*(2)} | Y^{*(1)} = 2, Y = 2) = 1$. The naive analysis yields low bias for $\boldsymbol{\beta}$ terms and for $\boldsymbol{\gamma^{(2)}_{111}}$ terms, but the rMSE is higher than that of the EM algorithm estimates, in particular (Table \ref{parameter-results-table-setting6-7}). The EM algorithm performs well in terms of bias and rMSE for most terms, but estimation is problematic for $\gamma^{(2)}_{1220}$ and  $\gamma^{(2)}_{122Z^{(2)}}$. These results are not concerning because extreme parameter estimates correspond to a lack of misclassification in the specificity mechanism of the second-stage outcomes, which was appropriate given the simulation design. The MCMC estimation did not show this extreme behavior in the estimates of $\gamma^{(2)}_{1220}$ and  $\gamma^{(2)}_{122Z^{(2)}}$ due to the limits of the uniform prior distribution used in the analysis. For other terms in the model, MCMC also produced reasonable estimates, though bias and rMSE were generally higher than that of the EM algorithm, especially for second-stage $\boldsymbol{\gamma^{(2)}}$ parameters. Both the EM algorithm and MCMC estimates correspond to highly accurate estimates of response and first-stage classification probabilities $P(Y = 1)$, $P(Y^{*(1)} = 1 | Y = 1)$, and $P(Y^{*(1)} = 2 | Y = 2)$ (Table \ref{probability-results-table-2}). The EM algorithm estimates $P(Y^{*(2)} = 1 | Y^{*(1)} = 1, Y = 1)$ without bias, while MCMC overestimates the quantity. Importantly, both proposed methods correctly estimate near-perfect classification for $P(Y^{*(2)} = 2 | Y^{*(1)} = 2, Y = 2)$, as specified by the simulation design.

\end{appendices}

\newpage

\nocite{*}
\bibliographystyle{apalike}
\bibliography{references}

\begin{thebibliography}{}

\bibitem[Angwin and Larson, 2016]{angwin2016Propublica}
Angwin, J. and Larson, J. (2016).
\newblock Machine {B}ias.
\newblock URL.
\newblock \url{https://www.propublica.org/article/machine-bias-risk-assessments-in-criminal-sentencing} [accessed 16 February 2023].

\bibitem[Bahl et~al., 2023]{bahl2023algorithms}
Bahl, U., Topaz, C.~M., Oberm{\"u}ller, L., Goldstein, S., and Sneirson, M. (2023).
\newblock Algorithms in judges' hands: Incarceration and inequity in {B}roward {C}ounty, {F}lorida.
\newblock {\em SocArXiv}.

\bibitem[Balagot et~al., 2019]{balagot2019homeless}
Balagot, C., Lemus, H., Hartrick, M., Kohler, T., and Lindsay, S.~P. (2019).
\newblock The homeless coordinated entry system: the vi-spdat and other predictors of establishing eligibility for services for single homeless adults.
\newblock {\em Journal of Social Distress and the Homeless}, 28(2):149--157.

\bibitem[Beesley and Mukherjee, 2022]{beesley2020statistical}
Beesley, L.~J. and Mukherjee, B. (2022).
\newblock Statistical inference for association studies using electronic health records: handling both selection bias and outcome misclassification.
\newblock {\em Biometrics}, 78(1):214--226.

\bibitem[Berk et~al., 2021]{berk2021fairness}
Berk, R., Heidari, H., Jabbari, S., Kearns, M., and Roth, A. (2021).
\newblock Fairness in criminal justice risk assessments: The state of the art.
\newblock {\em Sociological Methods \& Research}, 50(1):3--44.

\bibitem[Berrar, 2019]{BERRAR2019546}
Berrar, D. (2019).
\newblock Performance measures for binary classification.
\newblock In Ranganathan, S., Gribskov, M., Nakai, K., and Schönbach, C., editors, {\em Encyclopedia of Bioinformatics and Computational Biology}, pages 546--560. Academic Press, Oxford.

\bibitem[Betancourt, 2017]{betancourt2017identifying}
Betancourt, M. (2017).
\newblock Identifying {B}ayesian mixture models.
\newblock URL.
\newblock \url{https://betanalpha.github.io/assets/case_studies/identifying_mixture_models.html} [accessed 30 September 2022].

\bibitem[Cadigan and Lowenkamp, 2011]{cadigan2011implementing}
Cadigan, T.~P. and Lowenkamp, C.~T. (2011).
\newblock Implementing risk assessment in the federal pretrial services system.
\newblock {\em Fed. Probation}, 75:30.

\bibitem[Chouldechova et~al., 2018]{chouldechova2018case}
Chouldechova, A., Benavides-Prado, D., Fialko, O., and Vaithianathan, R. (2018).
\newblock A case study of algorithm-assisted decision making in child maltreatment hotline screening decisions.
\newblock In {\em Conference on Fairness, Accountability and Transparency}, pages 134--148. PMLR.

\bibitem[Cohen, 2018]{cohen2018predicting}
Cohen, T.~H. (2018).
\newblock Predicting sex offender recidivism: Using the federal post conviction risk assessment instrument to assess the likelihood of recidivism among federal sex offenders.
\newblock {\em Journal of Empirical Legal Studies}, 15(3):456--481.

\bibitem[Copp et~al., 2022]{copp2022pretrial}
Copp, J.~E., Casey, W., Blomberg, T.~G., and Pesta, G. (2022).
\newblock Pretrial risk assessment instruments in practice: The role of judicial discretion in pretrial reform.
\newblock {\em Criminology \& Public Policy}, 21(2):329--358.

\bibitem[Corbett-Davies et~al., 2017]{corbett2017algorithmic}
Corbett-Davies, S., Pierson, E., Feller, A., Goel, S., and Huq, A. (2017).
\newblock Algorithmic decision making and the cost of fairness.
\newblock In {\em Proceedings of the 23rd {ACM SIGKDD} international conference on knowledge discovery and data mining}, pages 797--806.

\bibitem[Coston et~al., 2021]{coston2021characterizing}
Coston, A., Rambachan, A., and Chouldechova, A. (2021).
\newblock Characterizing fairness over the set of good models under selective labels.
\newblock In {\em International Conference on Machine Learning}, pages 2144--2155. PMLR.

\bibitem[Dempster et~al., 1977]{dempster1977maximum}
Dempster, A.~P., Laird, N.~M., and Rubin, D.~B. (1977).
\newblock Maximum likelihood from incomplete data via the {EM} algorithm.
\newblock {\em Journal of the Royal Statistical Society: Series B (Methodological)}, 39(1):1--22.

\bibitem[Demuth, 2003]{demuth2003racial}
Demuth, S. (2003).
\newblock Racial and ethnic differences in pretrial release decisions and outcomes: A comparison of {H}ispanic, {B}lack, and {W}hite felony arrestees.
\newblock {\em Criminology}, 41(3):873--908.

\bibitem[Demuth and Steffensmeier, 2004]{demuth2004impact}
Demuth, S. and Steffensmeier, D. (2004).
\newblock The impact of gender and race-ethnicity in the pretrial release process.
\newblock {\em Social Problems}, 51(2):222--242.

\bibitem[Desmarais and Singh, 2013]{desmarais2013risk}
Desmarais, S. and Singh, J. (2013).
\newblock Risk assessment instruments validated and implemented in correctional settings in the {U}nited {S}tates.

\bibitem[Dobbie et~al., 2018]{dobbie2018effects}
Dobbie, W., Goldin, J., and Yang, C.~S. (2018).
\newblock The effects of pre-trial detention on conviction, future crime, and employment: Evidence from randomly assigned judges.
\newblock {\em American Economic Review}, 108(2):201--240.

\bibitem[Eubanks, 2018]{eubanks2018automating}
Eubanks, V. (2018).
\newblock {\em Automating {I}nequality: How {H}igh-{T}ech {T}ools {P}rofile, {P}olice, and {P}unish the poor}.
\newblock St. Martin's Press.

\bibitem[Fogliato et~al., 2020]{fogliato2020fairness}
Fogliato, R., Chouldechova, A., and G’Sell, M. (2020).
\newblock Fairness evaluation in presence of biased noisy labels.
\newblock In {\em International {C}onference on {A}rtificial {I}ntelligence and {S}tatistics}, pages 2325--2336. PMLR.

\bibitem[Grossman et~al., 2023]{grossman2023racial}
Grossman, J., Nyarko, J., and Goel, S. (2023).
\newblock Racial bias as a multi-stage, multi-actor problem: An analysis of pretrial detention.
\newblock {\em Journal of Empirical Legal Studies}, 20(1):86--133.

\bibitem[Hagan, 1975]{hagan1975social}
Hagan, J. (1975).
\newblock The social and legal construction of criminal justice: A study of the pre-sentencing process.
\newblock {\em Social Problems}, 22(5):620--637.

\bibitem[Imai et~al., 2023]{imai2023experimental}
Imai, K., Jiang, Z., Greiner, D.~J., Halen, R., and Shin, S. (2023).
\newblock Experimental evaluation of algorithm-assisted human decision-making: Application to pretrial public safety assessment.
\newblock {\em Journal of the Royal Statistical Society Series A: Statistics in Society}, 186(2):167--189.

\bibitem[Jung et~al., 2024]{jung2024mitigating}
Jung, J., Corbett-Davies, S., Gaebler, J.~D., Shroff, R., and Goel, S. (2024).
\newblock Mitigating included- and omitted-variable bias in estimates of disparate impact.

\bibitem[Katz and Spohn, 1995]{katz1995effect}
Katz, C.~M. and Spohn, C.~C. (1995).
\newblock The effect of race and gender on bail outcomes: A test of an interactive model.
\newblock {\em American Journal of Criminal Justice}, 19:161--184.

\bibitem[Kleinberg et~al., 2018]{kleinberg2018human}
Kleinberg, J., Lakkaraju, H., Leskovec, J., Ludwig, J., and Mullainathan, S. (2018).
\newblock Human decisions and machine predictions.
\newblock {\em The {Q}uarterly {J}ournal of {E}conomics}, 133(1):237--293.

\bibitem[Lattimore et~al., 2020]{lattimore2020prevalence}
Lattimore, P.~K., Tueller, S., Levin-Rector, A., and Witwer, A. (2020).
\newblock The prevalence of local criminal justice practices.
\newblock {\em Fed. Probation}, 84:28.

\bibitem[Leslie and Pope, 2017]{leslie2017unintended}
Leslie, E. and Pope, N.~G. (2017).
\newblock The unintended impact of pretrial detention on case outcomes: Evidence from new york city arraignments.
\newblock {\em The Journal of Law and Economics}, 60(3):529--557.

\bibitem[Mansournia et~al., 2018]{mansournia2018separation}
Mansournia, M.~A., Geroldinger, A., Greenland, S., and Heinze, G. (2018).
\newblock Separation in logistic regression: Causes, consequences, and control.
\newblock {\em American Journal of Epidemiology}, 187(4):864--870.

\bibitem[{Mapping Pretrial Injustice}, 2023]{HowManyJurisdictionsUseEachTool?_2023}
{Mapping Pretrial Injustice} (2023).
\newblock Mapping pretrial injustice: A community driven database.
\newblock \url{https://pretrialrisk.com/national-landscape/how-many-jurisdictions- use-each-tool/}.
\newblock Accessed: 2023-08-28.

\bibitem[Marlowe et~al., 2020]{marlowe2020employing}
Marlowe, D.~B., Ho, T., Carey, S.~M., and Chadick, C.~D. (2020).
\newblock Employing standardized risk assessment in pretrial release decisions: Association with criminal justice outcomes and racial equity.
\newblock {\em Law and Human Behavior}, 44(5):361.

\bibitem[Milgram et~al., 2015]{milgram2015pretrial}
Milgram, A. et~al. (2015).
\newblock Pretrial risk assessment: Improving public safety and fairness in pretrial decision-making, 27 fed.
\newblock {\em Federal {S}entencing {R}eporter}, 216:220.

\bibitem[Mitchell et~al., 2021]{mitchell2021algorithmic}
Mitchell, S., Potash, E., Barocas, S., D'Amour, A., and Lum, K. (2021).
\newblock Algorithmic fairness: Choices, assumptions, and definitions.
\newblock {\em Annual Review of Statistics and Its Application}, 8:141--163.

\bibitem[{Pretrial Justice Institute}, 2017]{TheStateofPretrialJusticeinAmerica_2017}
{Pretrial Justice Institute} (2017).
\newblock \url{https://www.prisonpolicy.org/scans/pji/the_state_of_pretrial_in_america_pji_2017.pdf}.
\newblock Accessed: 2023-08-28.

\bibitem[{Pretrial Justice Institute}, 2019]{ScanofPretrialPractices_2019}
{Pretrial Justice Institute} (2019).
\newblock \url{https://static1.squarespace.com/static/61d1eb9e51ae915258ce573f/t/61df3e19dc500a1e42344351/1642020381052/Scan+of+Pretrial+Practices.pdf}.
\newblock Accessed: 2023-08-28.

\bibitem[Prins, 2019]{prins2019criminogenic}
Prins, S.~J. (2019).
\newblock Criminogenic or criminalized? {T}esting an assumption for expanding criminogenic risk assessment.
\newblock {\em Law and {H}uman {B}ehavior}, 43(5):477.

\bibitem[{R Core Team}, 2021]{stats2021R}
{R Core Team} (2021).
\newblock {\em R: A Language and Environment for Statistical Computing}.
\newblock R Foundation for Statistical Computing, Vienna, Austria.

\bibitem[Sacks et~al., 2015]{sacks2015sentenced}
Sacks, M., Sainato, V.~A., and Ackerman, A.~R. (2015).
\newblock Sentenced to pretrial detention: A study of bail decisions and outcomes.
\newblock {\em American Journal of Criminal Justice}, 40:661--681.

\bibitem[Schlesinger, 2005]{schlesinger2005racial}
Schlesinger, T. (2005).
\newblock Racial and ethnic disparity in pretrial criminal processing.
\newblock {\em Justice Quarterly}, 22(2):170--192.

\bibitem[Schlesinger, 2008]{schlesinger2008cumulative}
Schlesinger, T. (2008).
\newblock The cumulative effects of racial disparities in criminal processing.
\newblock {\em The Advocate}, 13(3):22--34.

\bibitem[Schneider, 2020]{schneider2020locked}
Schneider, V. (2020).
\newblock Locked out by big data: how big data algorithms and machine learning may undermine housing justice.
\newblock {\em Columbia Human Rights Law Review}, 52:251.

\bibitem[Skeem et~al., 2020]{skeem2020impact}
Skeem, J., Scurich, N., and Monahan, J. (2020).
\newblock Impact of risk assessment on judges’ fairness in sentencing relatively poor defendants.
\newblock {\em Law and {H}uman {B}ehavior}, 44(1):51.

\bibitem[Sloan et~al., 2023]{sloan2023effect}
Sloan, C., Naufal, G., and Caspers, H. (2023).
\newblock The effect of risk assessment scores on judicial behavior and defendant outcomes.
\newblock {\em Journal of Human Resources}, 58:1--58.

\bibitem[Stephens, 2000]{stephens2000dealing}
Stephens, M. (2000).
\newblock Dealing with label switching in mixture models.
\newblock {\em Journal of the Royal Statistical Society: Series B (Statistical Methodology)}, 62(4):795--809.

\bibitem[Stevenson, 2018]{stevenson2018assessing}
Stevenson, M. (2018).
\newblock Assessing risk assessment in action.
\newblock {\em Minn. L. Rev.}, 103:303.

\bibitem[Stevenson and Doleac, 2022]{stevenson2022algorithmic}
Stevenson, M.~T. and Doleac, J.~L. (2022).
\newblock Algorithmic risk assessment in the hands of humans.
\newblock {\em Available at SSRN 3489440}.

\bibitem[Sutton, 2013]{sutton2013structural}
Sutton, J.~R. (2013).
\newblock Structural bias in the sentencing of felony defendants.
\newblock {\em Social Science Research}, 42(5):1207--1221.

\bibitem[Viljoen et~al., 2019]{viljoen2019impact}
Viljoen, J.~L., Jonnson, M.~R., Cochrane, D.~M., Vargen, L.~M., and Vincent, G.~M. (2019).
\newblock Impact of risk assessment instruments on rates of pretrial detention, postconviction placements, and release: A systematic review and meta-analysis.
\newblock {\em Law and Human Behavior}, 43(5):397.

\bibitem[Webb, 2023]{combo}
Webb, K. A.~H. (2023).
\newblock {\em COMBO: Correcting Misclassified Binary Outcomes in Association Studies}.
\newblock R package version 1.0.0.

\bibitem[Webb and Wells, 2023]{hochstedler2023statistical}
Webb, K. A.~H. and Wells, M.~T. (2023).
\newblock Statistical inference for association studies in the presence of binary outcome misclassification.
\newblock {\em arXiv preprint arXiv:2303.10215}.

\bibitem[Wooldredge, 2012]{wooldredge2012distinguishing}
Wooldredge, J. (2012).
\newblock Distinguishing race effects on pre-trial release and sentencing decisions.
\newblock {\em Justice Quarterly}, 29(1):41--75.

\bibitem[Wooldredge et~al., 2017]{wooldredge2017ecological}
Wooldredge, J., Frank, J., and Goulette, N. (2017).
\newblock Ecological contributors to disparities in bond amounts and pretrial detention.
\newblock {\em Crime \& Delinquency}, 63(13):1682--1711.

\bibitem[Wooldredge et~al., 2015]{wooldredge2015impact}
Wooldredge, J., Frank, J., Goulette, N., and Travis~III, L. (2015).
\newblock Is the impact of cumulative disadvantage on sentencing greater for black defendants?
\newblock {\em Criminology \& Public Policy}, 14(2):187--223.

\bibitem[Zeng, 2018]{zeng2018jail}
Zeng, Z. (2018).
\newblock Jail inmates in 2016.
\newblock {\em National {C}riminal {J}ustice}, 251210.

\end{thebibliography}

\end{document}